\documentclass[prd,twocolumn,showpacs,byrevtex]{revtex4}
\usepackage[dvips]{graphicx}
\usepackage{dcolumn}
\usepackage{amsmath}
\newcommand{\rd}{\mathrm{d}}
\newcommand{\C}{\mathcal{C}}
\newcommand{\G}{\mathcal{G}}
\newcommand{\D}{\mathcal{D}}
\newcommand{\K}{\mathcal{K}}
\newcommand{\bK}{\bar{\mathcal{K}}}
\begin{document}
%
%
%
\preprint{LAUR 00-2523}
\title[Resumming the large-$N$ approximation for $\ldots$]
   {Resumming the large-$N$ approximation
    for time evolving quantum systems}
\author{Bogdan Mihaila}
\thanks{Present address:
   Physics Division,
   Argonne National Laboratory,
   Argonne, IL 60439}
\email{bogdan.mihaila@unh.edu}
\author{John~F.~Dawson}
\email{john.dawson@unh.edu}
\homepage{http://www.theory.unh.edu/resum}
\affiliation{Department of Physics,\\
   University of New Hampshire, Durham, NH 03824}
\author{Fred Cooper}
\email{cooper@schwinger.lanl.gov}
\affiliation{Theoretical Division,\\
   Los Alamos National Laboratory, Los Alamos, NM 87545}
\date{\today}
\begin{abstract}
In this paper we discuss two methods of resumming the leading and next
to leading order in $1/N$ diagrams for the quartic $O(N)$ model. These
two approaches have the property that they preserve both boundedness
and positivity for expectation values of operators in our numerical
simulations.  These approximations can be understood either in terms
of a truncation to the infinitely coupled Schwinger-Dyson hierarchy of
equations, or by choosing a particular two-particle irreducible vacuum
energy graph in the effective action of the Cornwall-Jackiw-Tomboulis
formalism. We confine our discussion to the case of quantum mechanics
where the Lagrangian is
$L(x,\dot{x}) = (1/2) \sum_{i=1}^{N} \dot{x}_i^2 - (g/8N) \, [ \,
\sum_{i=1}^{N} x_i^2 - r_0^2 \, ]^{2}$.
The key to these approximations is to treat both the $x$ propagator
and the $x^2$ propagator on similar footing which leads to a theory
whose graphs have the same topology as QED with the $x^2$ propagator
playing the role of the photon.  The bare vertex approximation
is obtained by replacing the exact vertex function by the bare one in
the exact Schwinger-Dyson equations for the one and two point
functions.  The second approximation, which we call the dynamic Debye
screening approximation, makes the further approximation of
replacing the exact $x^2$ propagator by its value at leading order in
the $1/N$ expansion.  These two approximations are compared with exact
numerical simulations for the quantum roll problem.  The bare vertex
approximation captures the physics at large and modest $N$ better than
the dynamic Debye screening approximation.
\end{abstract}
\maketitle
%
%
\section{Introduction}
\label{sec:intro}

The need to understand quantum systems in real time in a quantum
field theoretic setting arose from attempts to understand various
early universe scenarios.  These scenarios are based on the
evolution of scalar fields either through their role as inflation
fields or as topological defect forming fields.  One would like
to understand the quantum evolution of these fields rather than
rely on unjustified treatments based on studying their classical
evolution.  The study of the ``slow rollover'' transition in an
upside down harmonic approximation by Guth and Pi\cite{ref:Guth}
was the first attempt to understand whether classical
approximations could be justified. However, one really needed to
go beyond the harmonic approximation to address the nonlinear
aspects of double well (and Mexican hat) potentials.  These
non-linear aspects effect production of topological defects as
well as the nature of the oscillation at the bottom of the well
which causes reheating.

Our ultimate goal is to be able to describe accurately over relevant
time periods the nonlinear aspects of quantum field theory evolutions.
Although in one-dimensional quantum mechanics, one can rely on a
numerical solution of the Schr\"odinger equation to understand the
time evolution of the system accurately over long time periods, in
field theory contexts the numerical solution of the functional
Schr\"odinger equation is presently beyond the reach of the largest
computers.  One important question is how to decrease the number of
degrees of freedom in a manner consistent with certain physical
requirements such as conservation of energy, preservation of positivity
and boundedness of expectation values.  Although this is guaranteed in
variational approximations, approximations based on various truncation
schemes, whether perturbative or non-perturbative in nature often fail
to preserve these physical requirements. For example, naive truncations
of the coupled Green functions equations beyond the truncation at
the two-point function level lead to secular behavior (unboundedness
at late times).  This is also true for the $1/N$ expansion which is
derivable from an effective action.  The second question is, after
guaranteeing these properties, how accurately have we described the
time evolution.

The simplest truncations of the field theory have been based on
gaussian variational methods\cite{ref:Hartree,ref:GV}, or the
related leading order in large-$N$ approximation (LOLN)
\cite{ref:LOLN,ref:ctpN}.  These two approximations can be shown
to be equivalent to a classical Hamiltonian dynamics for the
variational parameters (or equivalently the Green functions) which
leads to probability conservation at the quantum level so that the
results always lead to conserved energy, and positive and bounded
expectation values.  Unfortunately, hard scatterings which lead to
thermalization are ignored so that important physics is left
out.  The approximation also is numerically inaccurate after a
few oscillations in quantum mechanical applications, unless the
anharmonic coupling constant $g$ divided by the number of fields
$N$ in an $O(N)$ model is quite small.  In this paper we will be
comparing our methods of going beyond mean field theory (Hartree
or large-$N$) with exact numerical simulations of a quantum
mechanical $O(N)$ model.  In this way we can see how accurate the
approximations are as a function of $N$ as well as study
numerically if the approximation maintains the various physical
requirements we posit, such as boundedness and positive
definiteness of expectation values. The reason for using this
quantum mechanical model is that exact simulations can be done at
\emph{all} $N$, so that accuracy of the method as a function of
the parameter $1/N$ can be studied.  By restricting ourselves to
a quantum mechanics problem we unfortunately will not be able to
study questions of thermalization.  A complementary approach has
been undertaken by Aarts, Bonini, and Wetterich\cite{abw} where
they consider \emph{classical} 1+1 dimensional $\phi^4$ field
theory (for $N=1$). There one can look at some aspects of
classical thermalization (as long as one keeps a cutoff because
of the Raleigh-Jeans divergence) but one is restricted to low
values of $N$ so one cannot study the $N$ dependence of the
result.  Also one cannot study the \emph{quantum} aspects of the
problem.  In the above paper, Aarts \emph{et.al.}\ study a
truncation of the Green functions at the four-point level, which
is known to lead to unboundedness and secularity in quantum
mechanical (as well as classical) applications. It will be
interesting in the future to apply the approximations we are
using here to classical 1+1 dimensional $\phi^4$ to see if, and
how well, they describe the thermalization.

There are several ways of approaching the problem of thinning the
degrees of freedom of the quantum field theory.  One of the earliest
was based on making a variational approximation to the functional
Schr\"odinger equation. The variational approach has the advantage of
leading to a Hamiltonian dynamical system for the variational
parameters as well as to a density matrix which has positivity
properties.  Energy conservation and positivity and boundedness of
expectation values are automatically guaranteed.  However, even for
the simple problem of the quantum roll, the gaussian, or time
dependent Hartree approximation, studied by Cooper, Pi and
Stancioff\cite{ref:Hartree}, and improvements which are based on trial
wave functions of the form of a polynomial times a
gaussian\cite{ref:CC}, were found to be only accurate for relatively
short time periods (one or a few oscillations) when compared to the
exact numerical solution of the Schr\"odinger equation.  In quantum
mechanics, except for exceptional situations, the wave function in
multiwell situations gets very complicated very quickly and is not
easily described by a small number of variational parameters.

A second approach has been a direct $1/N$ expansion of the path
integral in the Schwinger-Keldysh-Bakshi-Mahanthappa
closed time path formalism\cite{ref:CTP}.  In this approach the
connected Green functions have the property that they start at
order $G_{2n} \propto 1/N^{n-1}$.  Thus if we retain only a
certain order in the expansion, there is a truncation in the
order of Green functions retained.  This approach was applied
recently to the quantum roll problem\cite{ref:paper1} and was
found to suffer from the secularity problem --- although the
short time behavior of the result was improved by including $1/N$
corrections, an exact reexpansion in terms of $1/N$ leads to
corrections in the Green functions that are of the form $\pm
t/N$ and so the individual corrections become unbounded as well
as non positive definite.  In this approach, although energy is
conserved, individual contributions are not positive definite and
unphysical behavior is found.

A third approach has been to consider the complete set of equal
time Green functions. These obey first order local equations in
time, as in the Schr\"odinger approach.  This approach has been
nicely systematized and an equation for the generating functional
obtained by Wetterich and collaborators in a series of papers
\cite{ref:EQT}. However, naive truncations of the equal time
Green function hierarchy again have the problem that although
there is a conserved energy, one cannot show that this truncation
(except at the two-point level) corresponds to a positive
definite probability so that expectation values are not
necessarily bounded or positive definite. Truncated at the two
point function level, this approach is identical to the Hartree
approximation.  However, simulations based on truncations
assuming 6th order or 8th order 1-PI graphs, could be set to
zero, were carried out for the $O(N)$ $x^4$ th oscillator problem,
and secularity was found for many choices of initial
conditions\cite{ref:bett}.  So we, as quantum field theorists,
having entered the domain of nonequilibrium phenomena, are now
beset with all the problems faced by our plasma and condensed
matter brethren more than 40 years ago!

In both quantum and classical many-body systems, the dynamical
equations are an infinite hierarchy of coupled equations which relate
given ensemble averages, whether classical or quantum, to successively
more complicated ones.  To make the solution of this hierarchy
possible, some truncation scheme is necessary.  Most naive truncation
schemes which, for example, just truncate the hierarchy of coupled
correlators at a particular order, do not preserve various physical
properties required of the system --- such as positivity of the
spectral components of the Green function and conservation of
probability.  A corollary of this is that in most perturbation
schemes, secularity arises quickly with each term in the perturbation
series, growing with higher powers of the time $t$.  In his seminal
paper of 1961, Robert Kraichnan\cite{ref:Robert} discussed in detail
the key issues and obtained a partial solution to the problem by
demanding that the approximations one should use should correspond to
some physically realizable dynamical system.  This would guarantee
positivity and secularity would be avoided.  The reason why
variational approximations avoid these problems is exactly because
they lead to a Hamiltonian dynamical system for the variational
parameters (which are related to equal time correlation functions).
He also discussed scenarios where particular classes of graphs, which
contained the relevant dynamics, are summed and he suggested some
physically motivated approximations which did not suffer from any
diseases.  In field theory one rarely has the parameter control to
make such guesses, however some progress in QCD has been made by
summing hard thermal loops\cite{ref:SDQCD}, which already tells us
some of the graphs that we want to include.  In plasma physics, one
wants to make sure that the approximation to the dynamics is robust
enough so that the photon propagator includes polarization effects,
which give Debye screening.  This is related to the hard thermal loop
summation in QCD.

To find resummation schemes that avoid the secularity problem we
will rely on the experience of our many-body and plasma physics
friends. To calculate the conductivity of a non-relativistic
plasma, it is known what graphs are necessary to sum in order to
get agreement with experimental
results\cite{ref:plasma1,ref:plasma2}.  Basically the
conductivity is found from the vertex function which must satisfy
an integral equation which sums ladders of the Debye screened
photon propagator.  The two approximations we will discuss here
will differ on whether the equivalent of the Debye screened
photon propagator for the anharmonic oscillator is treated in
lowest order in mean field theory, or is self-consistently
determined.  In studying the conductivity of a relativistic
plasma the first approximation has the advantage of obeying the
correct Ward identities (but violating energy conservation to
order $1/N$) whereas the second preserves energy conservation but
violates Ward identities (to order $1/N^2$).  Here we are not
studying QED, and the Ward identities of the $O(N)$ model for the
quantum mechanics problem are much simpler than those of QED and
energy conservation is a more important constraint on the
accuracy of the answer.  We will include both approximations here
mainly because of the recent interest in the gauge invariant
approximation for the relativistic plasma\cite{ref:emil}, and
also because in truncations of Schwinger-Dyson equations, it is
often too difficult to solve for the photon propagator self
consistently, and so one is often forced to try the more drastic
approximation of using the mean field propagator in the
resummation scheme. By studying this approximation in a quantum
mechanics problem we will see the shortcomings of such an
approach.

In what follows we will discuss two approaches to obtaining the
above two truncations of the exact Schwinger-Dyson equation and
apply them to the problem of the quantum roll --- the long time
behavior of $N$ coupled anharmonic oscillators with ``radial''
symmetry in an $N$-dimensional space.  This particular problem
has been studied by us previously\cite{ref:paper1} exactly and in
the next to leading order in the large-$N$ approximation (NLOLN)
and is interesting because exact numerical solutions can be found
for arbitrary $N$. What we found previously, is that for the
parameter set studied ($g \approx 1, M^2 = 2$), the next to
leading order in large-$N$ contributions became unbounded for $N
< 21$.  For larger $N$, where the approximation was physical, it
had the failing that it was unable to track the spreading of the
exact wave function which led to the envelope of the oscillations
found for $\langle \hat{x}^2(t) \rangle$ contracting at late
times and then reexpanding.  A related study of large-$N$ for
quantum mechanics in the context of the equal time correlators by
Bettencourt and Wetterich\cite{ref:bett}, also displayed growing
modes for various initial conditions.

The resummation presented here will allow one to track the
contraction for some period, but at later times it also fails in
that it leads to small oscillations about a fixed point value. In
field theory settings, where one hopes that this approximation
will lead to thermalization, optimistically this fixed point
behavior will become physical and be related to thermal
equilibration. Whether this is true or not can be checked by
studying this approximation for classical evolutions  averaged
over  a distribution  of initial conditions described by a
an initial  probability distribution in phase space.

In what follows we will present numerical solutions for the quantum
roll problem for the $O(N)$ model, and compare them to these two
different approximations to the Schwinger-Dyson equations, which sum
infinite numbers of leading order and next to leading order in $1/N$
graphs.  Our approach will be to introduce a composite ``field''
\begin{equation*}
  \chi
   = \frac{g}{2}
   \left  (
        \sum_{i=1}^{N} x_i^2 - r_0^2
   \right ) \>,
\end{equation*}
which is treated on equal footing to the field $x$.  By doing
that, the Schwinger-Dyson equations for the theory will have the
same topology as those of QED with $x$ playing the role of the
electron and $\chi$ the role of the photon.  At leading order in
large $N$ in N-flavor QED, one sums all the fermion loop vacuum
polarization corrections to the photon propagator which gives the
Debye screening. Here the bare photon propagator is replaced by a
local interaction in the graphs for the $\chi$ propagator in
LOLN.  The next consideration, important for charged plasmas, is
that to obtain reasonable agreement with experiments on the
conductivity of the plasma, the vertex function must sum all the
ladders with the Debye screened propagator as the kernel in the
integral equation.  The two resummation schemes which we discuss
in this paper both have this property.

The approximation which we call the bare vertex approximation (BVA),
uses the full Green function for $x$ as well as the full Green
function for $\chi$ in a 2-PI Hartree graph contribution to the
effective action.  This is in contrast to an earlier scheme for going
beyond $1/N$\cite{ref:Hu} using the 2-PI formalism which is based only
on the $x$ Green functions. The BVA approximation sums an infinite
Geometric series of 2-PI graphs of the single field formalism.  Recent
simulations in a toy 1+1 dimensional scalar field
theory\cite{ref:berges} show that the approximation described
in\cite{ref:Hu} already has the ability to thermalize arbitrary
initial conditions, so we are confident that the BVA approximation
will also have that feature when applied to a field theory problem.
The BVA can also be obtained by setting the full vertex function to
unity in the Schwinger-Dyson equations for the one- and two-point
functions with external sources hence the origin of its name.  The
second approximation we will study, which we call the dynamic Debye
screening approximation (DDSA), makes the further assumption that the
full $\chi$ propagator can be replaced by the lowest order in $1/N$
composite field propagator in all the integral equations.  The main
interest in the DDSA results from it being the lowest order
resummation scheme that \emph{exactly} preserves QED Ward identities.
Both these approximations are free from the difficulties found in the
perturbative $1/N$ expansion, which we display for comparison.  We
find that the BVA is accurate at modest times $\leq 25$ oscillations
when $N > 10$. At later times it settles down to oscillating about an
unphysical fixed point. The DDSA approximation violates energy
conservation at order $1/N$ and as a result becomes inaccurate after
several oscillations. In spite of this, it is numerically more
accurate for a longer period of time than the Hartree approximation at
small and modest values of $N$.

It should be kept in mind that quantum mechanics and quantum field
theory are very different.  For example, in the quantum mechanics
application discussed here, the graphs of the $O(1/N)$ corrections do
not correspond to interparticle collisions (as they do in field
theory) since we are restricting ourselves to one-particle quantum
mechanics. Nevertheless quantum mechanical examples provide excellent
test beds for key issues such as positivity violation, boundedness,
and late time accuracy of the approximations.  It is precisely these
questions that we are hoping to understand in this paper.

%
%
\section{The $O(N)$ model}
\label{sec:model}

The classical Lagrangian for the $O(N)$ model of $N$ non-linear
oscillators is given by:
\begin{equation}
   L(x,\dot{x})
   =
   \frac{1}{2} \sum_{i=1}^{N} \dot{x}_i^2 -
   \frac{g}{8N}
   \biggl (
      \sum_{i=1}^{N} x_i^2 - r_0^2
   \biggr )^{\!\!2}  \>.
   \label{SD.eq:classLi}
\end{equation}
The Schr\"odinger equation for this problem is given by:
\begin{equation}
   \left  \{
      - \frac{1}{2} \sum_{i=1}^N \frac{\partial^2}{\partial x_i^2}
      + V (r)
   \right \} \, \psi(x,t)
   =
   i \, \frac{\partial \psi(x,t)}{\partial t} \>,
\end{equation}
where $V(r)$ is a potential of the form
\begin{equation}
   V(r)  =  \frac{g}{8 N} \,
      \left  (
         r^2  -  r_0^2
      \right )^2
   \>,  \qquad
   r^2 = \sum_{i=1}^N x_i^2 \>.
   \label{eq:classi}
\end{equation}
For the quantum roll problem there is spherical symmetry.  This means
that we can assume a solution of the form $\psi(r,t) = \phi(r,t) /
r^{(N-1)/2}$, in which case the time dependent Schr\"odinger equation
for $\phi(r,t)$ reduces to\cite{ref:BlazotRipka}:
\begin{equation}
   \left \{
      - \frac{1}{2}
      \frac{\partial^2}{\partial r^2}
      + U(r)
   \right \} \, \phi(r,t)
   =
   i \, {\partial \phi(r,t) \over \partial t} \>,
   \label{eq:redham}
\end{equation}
with an effective one dimensional potential $U(r)$ given by
\begin{equation}
   U(r)
   =
   \frac{(N-1)(N-3)}{8 \, r^2}
   +
   \frac{g}{8N} \, \left ( r^2 - r_0^2 \right )^2
   \>.
\label{eq:Uofr}
\end{equation}
It is this equation that we will solve numerically to obtain exact
numerical solutions as a function of $N$.  $U(r)$ has a minimum at
$r=r_{\text{min}}$.  In our simulations, we have fixed our mass scale
$M^2$, defined as the second derivative of $U(r)$ at the minimum, to
have a value of 2, independent of $N$.

Returning to the Lagrangian formulation, it is useful for the purposes
of obtaining a large-$N$ expansion to introduce scaled variables:
\begin{alignat}{2}
   x_i &\rightarrow \sqrt{N} x_i \>,
   &\qquad
   r_0 &\rightarrow \sqrt{N} r_0 \>.
   \label{SD.eq:scaling}
\end{alignat}
Then the Lagrangian scales by a factor of $N$:
\begin{equation}
   L/N
   =
   L_{N}(x,\dot{x})
   =
   \frac{1}{2} \sum_{i=1}^{N} \dot{x}_i^2 -
   \frac{g}{8}
   \biggl (
      \sum_{i=1}^{N} x_i^2 - r_0^2
   \biggr )^{\!\!2}  \>.
   \label{SD.eq:classLii}
\end{equation}
We use these scaled variables in this paper, so that the rescaled $r_0
\approx 1$.  Next we introduce a composite coordinate $\chi$ by adding
to \eqref{SD.eq:classLii} a term:
\begin{equation}
   \frac{1}{2g}
   \biggl (
      \chi - \frac{g}{2} \Bigl (
                            \sum_{i=1}^{N} x_i^2 - r_0^2
                         \Bigr )
   \biggr )^{\!\!2}  \>.
   \label{SD.eq:addchipot}
\end{equation}
The Lagrangian \eqref{SD.eq:classLii} then becomes:
\begin{multline}
   L_{N}(x,\chi;\dot{x},\dot{\chi})
   =
   \sum_{i=1}^{N}
   \left  [
      \frac{1}{2} (
         \dot{x}_i^2 - \chi \, x_i^2
                  ) + j_i x_i
   \right ]
   \\
   + \frac{r_0^2 \, \chi}{2} + \frac{\chi^2}{2 g} + J \chi \>,
   \label{SD.eq:classLiii}
\end{multline}
where we have also added sources $j_i$ and $J$ coupling to $x_i$ and
$\chi$ respectively.  From this Lagrangian we get the Heisenberg
equations of motion for the operators $\hat{x}_i(t)$ and
$\hat\chi(t)$:
\begin{gather}
   \hat{\ddot{x}}_i(t) + \hat\chi(t) \, \hat{x}_i(t)
   =
   j_i(t) \>,
   \notag \\
   \frac{\hat\chi(t)}{g}
   =
   \frac{1}{2}
   \left  (
      \sum_{i=1}^{N} \hat{x}_i^2(t) - r_0^2
   \right ) - J(t) \>.
\end{gather}
Here, and in the following, we indicate operators by ``hats.''  Taking
expectation values with respect to an initial density matrix we obtain
the c-number equations:
\begin{gather}
   \langle \hat{\ddot{x}}_i(t) \rangle
   +
   \langle \hat\chi(t) \hat{x}_i(t) \rangle
   = j_i \>,
   \notag \\
   \frac{\langle \hat\chi(t) \rangle}{g}
   =
   \frac{1}{2}
   \left(
      \left\langle
         \sum_{i=1}^{N} \hat{x}_i^2(t)
      \right\rangle
      - r_0^2
   \right ) - J(t)  \>.
\label{eq:expect}
\end{gather}
By rewriting the quartic interaction in terms of the composite field
$\chi$, the induced interaction of the form $\chi x_i^2$ is
reminiscent of $N$ flavor QED with interaction $A_\mu \bar \psi_i
\gamma^\mu \psi_i$. The fact that these two theories have the same
topological structure will allow us to use the intuition gained in
classical plasmas to make appropriate approximations.

To simplify notation we include all independent coordinates in one
vector.  We define:
\begin{align}
   x_\alpha(t)
   &=
   [ \chi(t), x_1(t), x_2(t), \ldots , x_N(t)] \>,
   \notag \\
   j_\alpha(t)
   &=
   [\tilde{J}(t), j_1(t), j_2(t), \ldots , j_N(t)]  \>.
   \label{eq:xjextended}
\end{align}
for $\alpha = 0, 1, \ldots, N$, and where $\tilde{J}(t) = J(t) - r_0^2
/ 2$.  Absorbing the factor $r_0^2/2$ into the current means that
$\tilde{J}(t)$ is not zero when $J(t)$ is set to zero.  Greek indices
run from $0$ to $N$, whereas Latin indices go from $1$ to $N$.  Using
this extended notation, the generating functional $Z[j]$ and connected
generator $W[j]$ is given by the path integral:
\begin{equation}
   Z[j]
   =
   e^{i \, N W[j] }
   =
   \prod_{\alpha=0}^{N} \int \rd x_\alpha \,
   \exp
   \Bigl \{
      i \, N S_{N}[x;j]
   \Bigr \}
   \label{SD.eq:Z}
\end{equation}
where the action $S_{N}[x;j]$ is given by:
\begin{multline}
   S_{N}[x;j]
   =
   - \frac{1}{2} \sum_{\alpha,\beta}
     \int_{\C} \rd t \!\! \int_{\C} \rd t' \,
   x_\alpha(t) \, \Delta_{\alpha,\beta}^{-1}[x](t,t') \, x_\beta(t')
   \\
   +
   \sum_\alpha \int_{\C} \rd t \, x_\alpha(t) \, j_\alpha(t) \>,
   \label{eq:SNxj}
\end{multline}
and where $\Delta_{\alpha,\beta}^{-1}[x](t,t')$ is given by:
\begin{equation}
   \Delta_{\alpha,\beta}^{-1}[x](t,t')
   =
   \begin{pmatrix}
      D^{-1}(t,t')   & 0                  \\
      0              & G_{ij}^{-1}[\chi](t,t')
   \end{pmatrix}  \>,
   \label{SD.eq:Ginvdef}
\end{equation}
with
\begin{align}
   D^{-1}(t,t')
   &=
      - \frac{1}{g}
    \, \delta_{\C}(t,t') \>,
   \notag \\
   G_{ij}^{-1}[\chi](t,t')
   &=
   \biggl \{
      \frac{\rd^2}{\rd t^2}
      + \chi(t)
   \biggr \} \, \delta_{ij} \delta_{\C}(t,t') \>.
   \label{SD.eq:dginv}
\end{align}
In what follows it will be useful to introduce another matrix inverse
Green function $G^{-1}_{\alpha \beta}[x](t,t')$ as follows:
\begin{align}
   G_{\alpha,\beta}^{-1}[x](t,t')
   =
   - \frac{ \delta^2 S_N[x;j] }{\delta x_{\alpha}(t) \, \delta
   x_{\beta}(t') }
   \notag \\
   =
   \begin{pmatrix}
      D^{-1}(t,t')   & \bar{K}_j^{-1}(t,t') \\
      K_i^{-1}(t,t') & G_{i,j}^{-1}(t,t')
   \end{pmatrix}  \>,
   \label{SD.eq:ginvdef}
\end{align}
with $D^{-1}(t,t')$ and $G_{i,j}^{-1}(t,t')$ given by
Eq.~\eqref{SD.eq:dginv}, and $K_i^{-1}[x](t,t') = \bar{K}_i^{-1}[x](t,t')
= x_i(t) \, \delta_{\C}(t,t')$.

%
%
\section{The Schwinger-Dyson equations}
\label{sec:SD}

The Schwinger-Dyson equations are integral equations for the Green
functions.  The Green functions can be obtained by functional
differentiation of the path integral for the generating function
in the presence of external sources.  After setting the external
sources to zero, one obtains an infinitely coupled hierarchy of
coupled equations for the Green functions.  For an initial value
problem, the boundary conditions on the Green functions can be
implemented by using a time ordered product where the time
ordering refers to the closed time path contour of the
Schwinger-Keldysh-Bakshi-Mahanthappa formalism\cite{ref:CTP}.  A
detailed discussion of that formalism as applied to implementing
the $1/N$ expansion for this particular problem is described in
ref.~\cite{ref:ctpN}.  One way to generate the equations is to
consider the identity\cite{ref:IZ}:
\begin{equation}
   \prod_\beta \int \rd x_\beta
   \> \frac{\delta }{\delta x_\alpha(t)} \> e^{iN \, S_{N}[x;j] }
   = 0 \>,
   \label{SD.eq:identity}
\end{equation}
from which we find:
\begin{gather}
   - \frac{1}{g} \, \chi(t)
   +
   \frac{1}{2}
   \biggl \{
      \sum_i
      \biggl [
         x_i^2(t)
         +
         \frac{1}{N} \mathcal{G}_{ii}(t,t)/i
      \biggr ]
      -
      r_0^2
   \biggr \}
   =
   J(t)  \>,
   \notag \\
   \biggl \{
      \frac{\rd^2}{\rd t^2}
      + \chi(t)
   \biggr \} \, x_i(t)
   +
   \frac{1}{N} \mathcal{K}_i(t,t) / i
   =
   j_i(t) \>,
   \label{SD.eq:xchieq}
\end{gather}
where $x_i(t)$ and $\chi(t)$ are \emph{average values} of the
operators,
\begin{align*}
   x_i(t)
   &\equiv
   \frac{\delta  W[J,j]/i }{\delta j_i(t)}
   =
   \langle \hat{x}_i(t) \rangle \>,
   \\
   \chi(t)
   &\equiv
   \frac{\delta  W[J,j]/i}{\delta J(t)}
   = \langle \hat\chi(t) \rangle \>,
\end{align*}
and where the Green functions $\mathcal{G}_{\alpha,\beta}[j](t,t')$ are
defined by:
\begin{align}
   \mathcal{G}_{\alpha,\beta}[j](t,t')
   &=
   \frac{\delta x_\alpha(t)}{\delta j_\beta(t')}
   =
   \frac{\delta^2 W[j]}{\delta j_\alpha(t) \, \delta j_\beta(t')}
   \notag \\
   &=
   \begin{pmatrix}
      \mathcal{D}(t,t')       & \mathcal{K}_j(t,t') \\
      \bK_i(t,t') & \mathcal{G}_{i,j}(t,t')
   \end{pmatrix}  \>.
   \label{SD.eq:GGdef}
\end{align}
Eq.~\eqref{SD.eq:xchieq} is identical to Eq.~\eqref{eq:expect}.
In this equation and in what follows, $x_i$ and $\chi$ now
correspond to the expectation values:

The Green functions are explicitly given by
\begin{alignat*}{2}
   \mathcal{D}(t,t')
   &=
   \frac{\delta^2 W[J,j]}{\delta J(t) \, \delta J(t')}
   & \qquad
   \mathcal{K}_i(t,t')
   &=
   \frac{\delta^2 W[J,j]}{\delta J(t) \, \delta j_i(t')}
   \\
   \bK_i(t,t')
   &=
   \frac{\delta^2 W[J,j]}{\delta j_i(t) \, \delta J(t')}
   & \qquad
   \mathcal{G}_{i,j}(t,t')
   &=
   \frac{\delta^2 W[J,j]}{\delta j_i(t) \, \delta j_j(t')} \>.
   \label{SD.eq:GGmatdef}
\end{alignat*}
The integrability conditions require that $\bK_i(t,t') =
\mathcal{K}_i(t',t)$.  To obtain the Schwinger-Dyson equations it is
advantageous to Legendre transform to the expectation value of the
coordinate variables $x_\alpha(t)$, as the independent variable
instead of the currents.  The effective action generating functional
of 1-PI graphs is given by a Legendre transformation:
\begin{equation}
   \Gamma[x]
   =
   W[j]
   -
   \int_{\C} \rd t \, \sum_\alpha \{ x_\alpha(t) j_\alpha(t) \} \>.
\end{equation}
So since $j_\alpha(t) = - \delta \Gamma[x] / \delta x_\alpha(t)$, the
equations of motion \eqref{SD.eq:xchieq} give values for derivatives
of $\Gamma[x]$:
\begin{align}
   - \frac{\delta \Gamma[x]}{\delta \chi(t)}
   &=
   - \frac{1}{g} \, \chi(t)
   \notag \\ & \>
   +
   \frac{1}{2}
   \biggl \{
      \sum_i
      \biggl [
         x_i^2(t)
         +
         \frac{1}{N} \mathcal{G}_{ii}(t,t)/i
      \biggr ]
      -
      r_0^2
   \biggr \}
   \label{SD.eq:Gammachieq} \\
   - \frac{\delta \Gamma[x]}{\delta x_i(t)}
   &=
   \biggl \{
      \frac{\rd^2}{\rd t^2}
      + \chi(t)
   \biggr \} \, x_i(t)
   +
   \frac{1}{N} \mathcal{K}_i(t,t) / i \>.
   \label{SD.eq:Gammaxeq}
\end{align}
However the Green functions here, $\mathcal{G}_{ii}(t,t)$ and
$\mathcal{K}_i(t,t)$ are defined in Eq.~\eqref{SD.eq:GGdef} as
functionals of the currents $j_\alpha(t)$.  These must be expressed as
functionals of $x_\alpha(t)$ by inverse relations.  We define these
inverse Green functions, which are functionals of $x_\alpha(t)$, by:
\begin{align*}
   \mathcal{G}_{\alpha,\beta}^{-1}[x](t,t')
   &=
   \frac{\delta j_\alpha(t)}{\delta x_\beta(t')}
   =
   - \frac{\delta^2 \Gamma[x]}{\delta x_\alpha(t) \, \delta
   x_\beta(t')}
   \\
   &=
   \begin{pmatrix}
      \mathcal{D}^{-1}(t,t')   & \bK_j^{-1}(t,t') \\
      \mathcal{K}_i^{-1}(t,t') & \mathcal{G}_{i,j}^{-1}(t,t')
   \end{pmatrix}  \>,
\end{align*}
where explicitly
\begin{align*}
   \mathcal{D}^{-1}(t,t')
   &=
   -\frac{\delta^2 \Gamma[\chi,x]}
         {\delta \chi(t) \, \delta \chi(t')} \>,
   \\
   \bK_i^{-1}(t,t')
   &=
   -\frac{\delta^2 \Gamma[\chi,x]}
         {\delta \chi(t) \, \delta x_i(t')} \>,
   \\
   \mathcal{K}_i^{-1}(t,t')
   &=
   -\frac{\delta^2 \Gamma[\chi,x]}
         {\delta x_i(t) \, \delta \chi(t')} \>,
   \\ 
   \mathcal{G}_{i,j}^{-1}(t,t')
   &=
   -\frac{\delta^2 \Gamma[\chi,x]}
         {
\delta x_i(t) \, \delta x_j(t')} \>.
\end{align*}
Again we have $\bK_i^{-1}(t,t') = \mathcal{K}_i^{-1}(t',t)$.  The
inverse Green functions are given by differentiating the equations
of motion, Eqs.~\eqref{SD.eq:Gammachieq} and \eqref{SD.eq:Gammaxeq},
with respect to the coordinates.  Using
\begin{equation*}
   \int_{\C} \rd t' \sum_\beta
   \mathcal{G}_{\alpha,\beta}^{-1}[x](t,t') \,
   \mathcal{G}_{\beta,\gamma}[j](t',t'')
   =
   \delta_{\alpha,\gamma} \delta_{\C}(t,t'') \>,
\end{equation*}
we find:
\begin{align}
   \frac{\delta \mathcal{G}_{\alpha,\beta}[j](t_1,t_2)}
        {\delta x_\gamma(t_3)}
   &=
   -
   \int_{\C} \rd t_4 \!\! \int_{\C} \rd t_5 \sum_{\delta,\epsilon}
   \mathcal{G}_{\alpha,\delta}[j](t_1,t_4) 
   \notag \\
   & \quad \times 
   \Gamma_{\delta,\epsilon,\gamma}[x](t_4,t_5,t_3) \,
   \mathcal{G}_{\epsilon,\beta}[j](t_5,t_2) \>,
   \label{SD.B.eq:dGdx}
\end{align}
where $\Gamma_{\alpha,\beta,\gamma}[x](t_1,t_2,t_3)$ is the three-point
vertex function, defined by:
\begin{align}
   \Gamma_{\alpha,\beta,\gamma}[x](t_1,t_2,t_3)
   &=
   \frac{\delta \mathcal{G}_{\alpha,\beta}^{-1}[x](t_1,t_2)}
        {\delta x_\gamma(t_3)}
   \notag \\
   &=
   - \frac{\delta^3 \Gamma[x]}
          {\delta x_\alpha(t_1) \, \delta x_\beta(t_2) \,
           \delta x_\gamma(t_3)} \>.
   \label{SD.eq:Gamma3def}
\end{align}
Explicitly, we find an equation of the form:
\begin{equation}
   \mathcal{G}_{\alpha,\beta}^{-1}(t,t')
   =
   G_{\alpha,\beta}^{-1}(t,t')
   +
   \Sigma_{\alpha,\beta}(t,t')  \>,
   \label{eq:GGinvGinvSigma}
\end{equation}
where $G_{\alpha,\beta}^{-1}(t,t')$ is given by
Eq.~\eqref{SD.eq:ginvdef}.  The generalized self energy
$\Sigma_{\alpha,\beta}(t,t')$ is given by:
\begin{equation}
   \Sigma_{\alpha,\beta}(t,t')
   =
   \begin{pmatrix}
      \Pi(t,t')          & \Omega_j(t,t') \\
      \bar\Omega_i(t,t') & \Sigma_{ij}(t,t')
   \end{pmatrix}  \>,
   \label{eq:Sigmasdefs}
\end{equation}
and where the polarization $\Pi(t,t')$, self energy
$\Sigma_{ij}(t,t')$, and the off diagonal terms $\Omega_i(t,t')$ and
$\bar\Omega_i(t,t')$ are given by:
\begin{widetext}
\begin{align}
   \Pi(t,t')
   &=
   \frac{i}{2N}
   \sum_{i,\alpha,\beta} \int_{\C} \rd t_1 \!\! \int_{\C} \rd t_2 \,
   \mathcal{G}_{i,\alpha}(t,t_1)
   \Gamma_{\alpha,\beta,0}(t_1,t_2,t')
   \mathcal{G}_{\beta,i}(t_2,t) \>,
   \notag \\
   \Sigma_{ij}(t,t')
   &=
   \frac{i}{N}
   \sum_{\alpha,\beta} \int_{\C} \rd t_1 \!\! \int_{\C} \rd t_2 \,
   \mathcal{G}_{i,\alpha}(t,t_1) \,
   \Gamma_{\alpha,\beta,j}(t_1,t_2,t')
   \mathcal{G}_{\beta,0}(t_2,t) \,
   \notag \\
   \Omega_i(t,t')
   &=
   \frac{i}{2N}
   \sum_{j,\alpha,\beta} \int_{\C} \rd t_1 \!\! \int_{\C} \rd t_2 \,
   \mathcal{G}_{j,\alpha}(t,t_1)
   \Gamma_{\alpha,\beta,i}(t_1,t_2,t')
   \mathcal{G}_{\beta,j}(t_2,t)
   \notag \\
   \bar\Omega_i(t,t')
   &=
   \frac{i}{N}
   \sum_{\alpha\beta} \int_{\C} \rd t_1 \!\! \int_{\C} \rd t_2 \,
   \mathcal{G}_{i,\alpha}(t,t_1)
   \Gamma_{\alpha,\beta,0}(t_1,t_2,t')
   \mathcal{G}_{\beta,0}(t_2,t) \>.
   \label{eq:allSigmas}
\end{align}
\end{widetext}
In order to solve the equation for the two point function,
Eq.~\eqref{eq:GGinvGinvSigma}, one requires knowledge of the three
point function, defined by Eq.~\eqref{SD.eq:Gamma3def}.  This in turn
requires knowledge of the four-point function, \emph{ad infinitum}.
It is this infinite hierarchy of coupled Green function equations
that corresponds to solving exactly the Schr\"odinger equation.

The matrix inversion of Eq.~\eqref{eq:GGinvGinvSigma} gives the set
of coupled equations,
\begin{multline}
   \mathcal{G}_{\alpha,\beta}(t,t')
   =
   G_{\alpha,\beta}(t,t')
   -
   \sum_{\gamma,\delta} \int_{\C} \rd t_1 \int_{\C} \rd t_2 \,
   G_{\alpha,\gamma}(t,t_1) 
   \\ \times 
   \Sigma_{\gamma,\delta}(t_1,t_2) \,
   \mathcal{G}_{\delta,\beta}(t_2,t')  \>,
   \label{SD.eq:Gmateq}
\end{multline}
where
\begin{equation}
   G_{\alpha,\beta}(t,t')
   =
   \begin{pmatrix}
      D(t,t')         & K_i(t,t')          \\
      \bar{K}_i(t,t') & G_{ij}(t,t')
   \end{pmatrix}  \>.
   \label{SD.eq:Gdef}
\end{equation}
with
\begin{gather}
   \begin{split}
   \sum_j
   \biggl \{
      \biggl [
         \frac{\rd^2}{\rd t^2} + \chi(t)
      \biggr ] \, \delta_{ij}
      &+
      g \, x_i(t) \, x_j(t)
   \biggr \} \, G_{jk}(t,t')
   \label{eq:defGij} \\ & 
   = \delta_{ik} \, \delta_{\C}(t,t')  \>,
   \end{split} \\
   D(t,t')
   =
   - g \, \delta_{\C}(t,t') +
   g^2 \sum_{ij} x_i(t) \, G_{ij}(t,t') \, x_j(t') \>,
   \label{eq:defD} \\
   \bar{K}_j(t,t')
   =
   K_j(t',t)
   =
   g \sum_i G_{ji}(t,t') \, x_i(t') \>.
   \label{eq:defKbarK}
\end{gather}
When $x_i(t) \neq 0$, one notes that $D(t,t')$ is not the inverse of
$D^{-1}(t,t')$.

The vertex function $\Gamma_{\alpha,\beta,\gamma}[x](t_1,t_2,t_3)$
defined in \eqref{SD.eq:Gamma3def} is obtained by differentiation
of Eq.~\eqref{eq:GGinvGinvSigma} with respect to $x_\gamma(t)$.  We
find:
\begin{multline}
   \Gamma_{\alpha,\beta,\gamma}[x](t_1,t_2,t_3)
   =
   \frac{\delta \mathcal{G}_{\alpha,\beta}^{-1}[x](t_1,t_2)}
        {\delta x_\gamma(t_3)}
   \\
   =
   f_{\alpha,\beta,\gamma} \,
   \delta_{\C}(t_1,t_2) \, \delta_{\C}(t_1,t_3)
   +
   \Phi_{\alpha,\beta,\gamma}[x](t_1,t_2,t_3) \>.
   \label{SD.eq:threept}
\end{multline}
Here $f_{i,j,0} = f_{0,i,j} = f_{i,0,j} = \delta_{ij}$, otherwise $f$
is zero.  $\Phi_{\alpha,\beta,\gamma}[x](t_1,t_2,t_3)$ is given by
derivatives of the self-energy matrix:
\begin{equation}
   \Phi_{\alpha,\beta,\gamma}[x](t_1,t_2,t_3)
   =
   \frac{\delta \, \Sigma_{\alpha,\beta}[x](t_1,t_2)}
        {\delta x_\gamma(t_3)}  \>,
   \label{SD.eq:Phidef}
\end{equation}
and is of order $1/N$.

We are interested in resummation schemes that are exact to order $1/N$
for $\langle x_i^2 \rangle$.  We see from Eqs.~\eqref{SD.eq:threept}
and \eqref{SD.eq:Phidef} that it is consistent to replace
$\Gamma_{\alpha,\beta,\gamma}[x](t_1,t_2,t_3)$ in
Eq.~\eqref{SD.eq:Gmateq} by the first term in
Eq.~\eqref{SD.eq:threept} to obtain a resummation which is exact to
order $1/N$.  To simplify our discussion of the exact Schwinger-Dyson
equation for the vertex function, we will only consider the case of
the quantum roll where $x_i(t)=0$.

Following the treatment of ref.~\cite{ref:CGHT}, we have for the
3-$\chi$ vertex:
\begin{widetext}
\begin{equation*}
   \Lambda(t_1,t_2,t_3)
   =
   \frac{\delta {\mathcal D}^{-1}(t_1, t_2)}{\delta \chi(t_3)}
   =
   \sum_{ijk} \int_{\C} \rd t_4 \!\! \int_{\C} \rd t_5 \>
      \mathcal{G}_{ij}(t_3,t_4) \,
      \mathcal{G}_{ik}(t_3,t_5) \,
      M_{jk}(t_4, t_2; t_5, t_1) \>,
\end{equation*}
where $M_{jk}(t_4, t_2; t_5, t_1)$ is 1-PI in the channel $x + x
\rightarrow \chi + \chi$.  The lowest order in $1/N$ contribution to
$M(t_4, t_5; t_2,t_3)$ is:
\begin{equation}
   M_{jk}(t_4, t_2; t_5, t_1)
   =
   \delta_{\C}(t_4, t_2) \delta_{\C}(t_5, t_1) \,
   \mathcal{G}_{jk}(t_2, t_1) \>.
\end{equation}
When $x_i(t) = 0$, the exact Schwinger-Dyson equation for the
$\chi$-$x$-$x$ vertex is
\begin{multline}
   \Gamma_{ij}(t_1,t_2,t_3)
   =
   \frac{\delta \mathcal{G}^{-1}(t_1,t_2)}{\delta \chi(t_3)}
   =
   \delta_{ij} \, \delta_{\C}(t_1,t_2) \, \delta_{\C}(t_1,t_3)
   -
   \int_{\C} \rd t_4 \!\! \int_{\C} \rd t_5 \!\!
   \int_{\C} \rd t_6 \!\! \int_{\C} \rd t_7 \, \times
   \\
   \Biggl \{
      \sum_{klmn}
      \Gamma_{kl}(t_4,t_5,t_3) \,
      \G_{km}(t_4,t_6) \,
      \G_{ln}(t_5,t_7) \,
      \K_{1\, mn}(t_5,t_2;t_7,t_1)
      +
      \Lambda(t_4,t_5,t_3) \,
      \D(t_4,t_5) \,
      \D(t_6,t_7) \,
      \K_{2\, ij}(t_5,t_2;t_7,t_1)
   \Biggr \}  \>.
   \label{eq:fullGamma}
\end{multline}
\end{widetext}
where $\K_1$ and $\K_2$ are the s-channel 2-PI scattering amplitudes
for the reactions: $x + x \rightarrow x + x$ and $\chi + \chi
\rightarrow x + x$, respectively.

\begin{figure*}[t]
   \centering
   \includegraphics[width=5.0in]{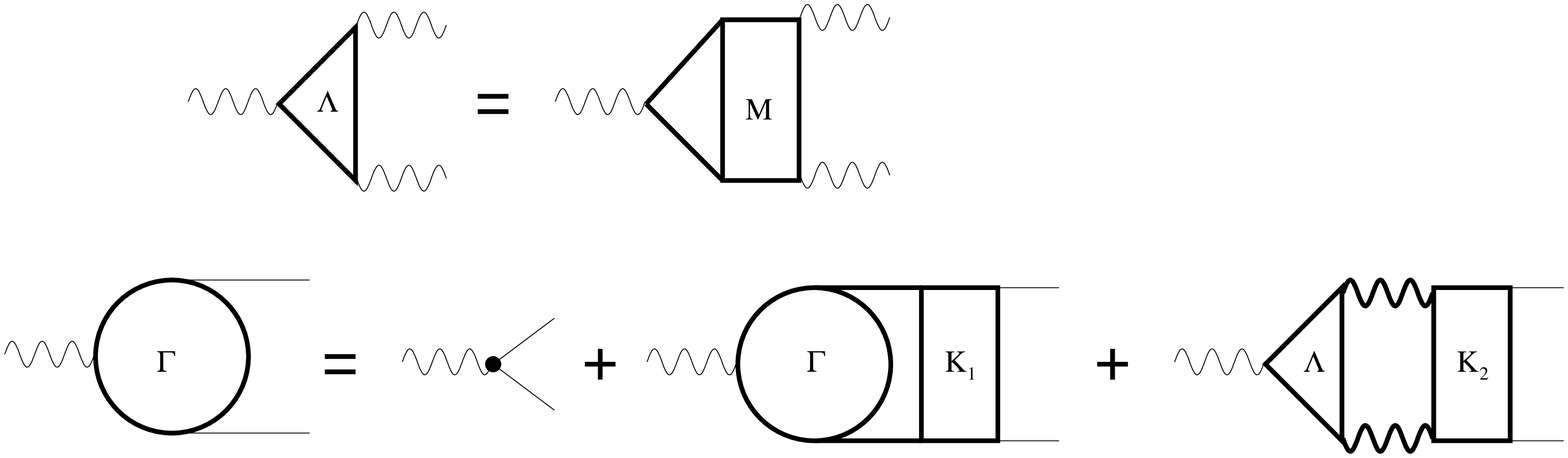}
   \caption{Schwinger-Dyson equations for the vertex function.
   Solid lines represent the
   $\mathcal{G}_{ij}(t,t')$ propagator and heavy wiggly lines are the
   $\mathcal{D}(t,t')$ propagator.}
   \label{fig:SDV}
\end{figure*}

This is shown pictorially in Fig.~\ref{fig:SDV}.  In general one
then has to obtain equations for the 2-PI scattering amplitudes as
well as for $\Lambda$. These will depend on even higher $n$-point
functions, \emph{ad infinitum}.  In our approximations made at the
two-point function level, the 2-PI s-channel scattering amplitudes
$K_1$ and $K_2$, used in the equations for the vertex function, will
turn out to be graphs for one-particle exchange in the t-channel of
the $\chi$- and $x$-particles respectively.

%
%
%
%

In our truncations of the Schwinger-Dyson equations, we will
always replace the full three-point vertex function by the bare
one in the equations for $x$ and $\mathcal{G}$ in the
\emph{presence} of external sources.  Once this truncation is
made, then for the problem we are addressing here (the
approximate time evolution of $N$ quantum anharmonic oscillators)
one never needs any of the $N$ point functions beyond the 1 and 2
point function equations.  What will distinguish a further
approximation we will call the DDSA is that we will also further
approximate the $\chi$ propagator to be that of the LOLN
approximation.

By making this bare vertex approximation in the equations for the one-
and two-point Green functions, we have \emph{not} relinquished our
ability to calculate in this approximation all the higher connected
Green functions.  These are obtainable by further functional
differentiation of the effective action.  In particular if we wanted
to use linear response theory (the Kubo formula) to obtain the
electrical conductivity for a QED plasma, one would functionally
differentiate the equation for the inverse two-point for the electron
function with respect to $A_\mu$.  In our problem the photon is
replaced by the composite field $\chi$, and the electron by $x_i$.

Because of recent interest in studying plasma conductivity in
both QED and QCD, we will spend extra time on comparing the
equations obtained for the vertex function in the three
approximations considered here. In conductivity calculations, it
is necessary to sum all the ladder graphs in the equation for the
vertex function to get good results for dilute plasmas.  We will
find that in NLOLN the vertex function is \emph{not} an integral
equation but is rather the sum of a few diagrams whereas the
other two approximations lead to integral equations that sum an
infinite number of diagrams.  Another issue is in preserving Ward
identities.  One of the reasons the large-$N$ expansion was so
interesting is that it is a complete reexpansion of the field
theory which preserves Ward identities at each order.  The QED
plasma conductivity problem people \cite{ref:emil} became
interested in the DDSA because it \emph{exactly} obeyed the Ward
identities, whereas the BVA approximation violates Ward
identities at order $1/N^2$.  It is for this reasons we thought
it appropriate to study the DDSA approximation, even though it
violated energy conservation already at order $1/N$, hoping that
at least at large $N$ it would be numerically accurate \emph{and}
satisfy Ward identities in QED applications.

The exact formula for the energy is given by:
\begin{widetext}
\begin{equation}
   E/N
   =
   \frac{1}{2}
   \left  \langle
      \sum_i
      \left  \{
         \hat{\dot{x}}_i^2(t) + \hat\chi(t) \, \hat{x}_i^2(t)
            - r_0^2 \hat\chi(t) - \hat\chi^2(t) / g
      \right \}
   \right \rangle  \>.
\end{equation}
When $x_i(t) =\langle \hat{x}_i(t) \rangle = 0$
and $\dot{x}_i(t) = 0$, one obtains:
\begin{align}
   E/N
   &=
   \frac{1}{2}
      \sum_i
      \biggl\{
         \left  .
             \frac{ \partial^2 \, \G(t,t')/i }
                  { \partial t \, \partial t' }
         \right |_{t'=t}
         +
         \chi(t) \, \G(t,t)/i
         -
         r_0^2 \chi(t)
         -
         \frac{1}{g} 
         \left  [ 
            \chi^2(t) + \frac{1}{N} \D(t,t)/i 
         \right ]
    \notag \\
    & \qquad
         +
         \frac{1}{N} \sum_{ijk}
         \int_{\C} \rd t_1 \!\! \int_{\C} \rd t_2 \!\!
         \int_{\C} \rd t_3 \,
         \D(t_1,t_2) \, \G_{ij}(t_1,t) \, \G_{ik}(t,t_1) \,
         \Gamma_{jk}(t_1,t_3,t_2)
    \biggr\} \>,
    \label{eq:fullenergy}
\end{align}
\end{widetext}
where $\Gamma_{jk}(t_1,t_3,t_2)$ is the full vertex function
given in Eq.~\eqref{eq:fullGamma}.
%
%
\section{Effective action for two-particle irreducible graphs}
\label{sec:effaction}

Since the approximations we are going to consider have a simple
interpretation in terms of keeping a particular 2-PI vacuum graph in
the generating functional of the 2-PI graphs, we would like to review
this formalism following the approach of Cornwall, Jackiw, and
Tomboulis (CJT)\cite{ref:CJT}.

The first Legendre transform of the generating functional $W[j]$ of
connected Green functions is widely known and used and is called the
``effective action.''  The higher Legendre transforms (second, third,
etc.) were introduced by De~Dominicis and Martin\cite{ref:DM} in
quantum statistics.  Dahmen and Jona-Lasinio\cite{ref:DJL}, and later
Visil'ev and Kazanskii\cite{ref:VK}, extended these ideas to quantum
field theory.  These methods were then used by Cornwall, Jackiw, and
Tomboulis to discuss dynamical symmetry breaking in Hartree type
approximations which later led to the second Legendre transformation
formalism being called the CJT formalism.  These higher order Legendre
transformed actions have the advantage of being able to treat higher
order Green functions on the same footing as the coordinates.

We will first summarize the general results of that paper before
proceeding to the specific approximations we consider in this paper.
The method of CJT is to introduce one- and two-body sources for the
coordinates $x_\alpha(t)$ and the Green functions
$\mathcal{G}_{\alpha,\beta}(t,t')$ in the action, and then make a
Legendre transformation to the one- and two-point functions.  The
resulting action, as a function of $x$ and $\mathcal{G}$, contains a
term which is the sum of all two-particle irreducible vacuum graphs.
This term can be written using the vertices of the interaction and
$\mathcal{G}$.  We use the extended notation for the coordinates and
one-body sources, given in Eq.~\eqref{eq:xjextended}.

%
%
\begin{figure*}[t]
   \centering
   \includegraphics[width=5.0in]{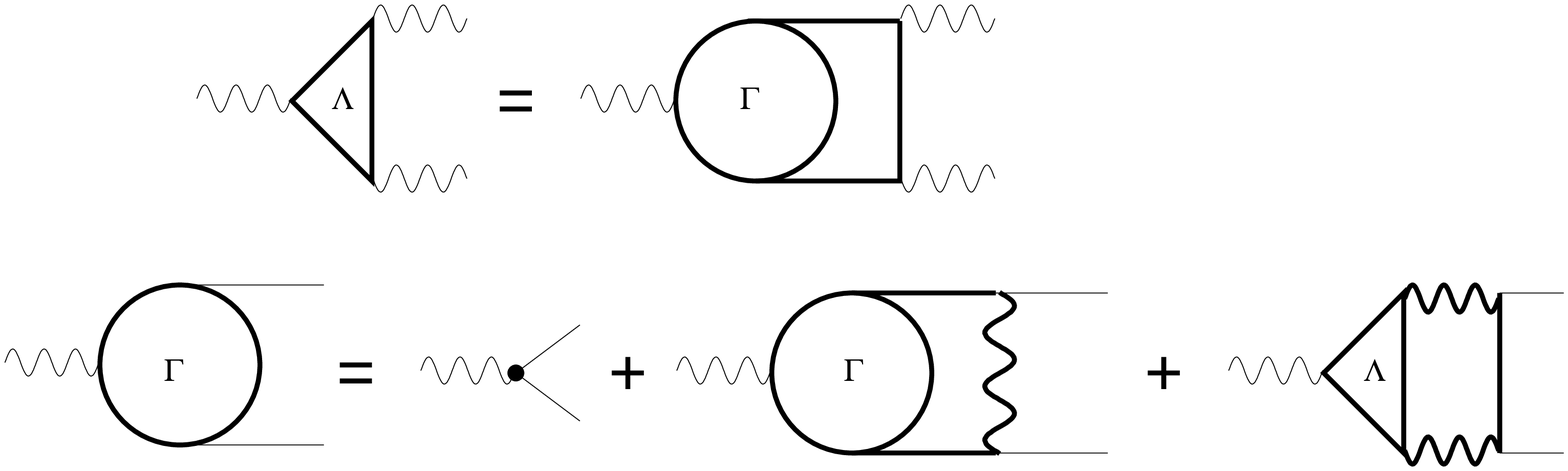}
   \caption{The vertex function for the BVA.  The top figure represents
   Eq.~\eqref{SD.eq:verteqa} and the bottom figure represents
   Eq.~\eqref{SD.eq:verteqb}.  Solid lines represent the
   $\mathcal{G}_{ij}(t,t')$ propagator and heavy wiggly lines are the
   $\mathcal{D}(t,t')$ propagator.}
   \label{fig:BVA}
\end{figure*}
%
%

Thus, the generating functional $Z[j,k]$ for the CJT action is given
by:
\begin{equation*}
   Z[j,k]
   =
   e^{ i N W[j,k] }
   =
   \prod_{\alpha=0}^{N} \int \rd x_{\alpha} \,
      \exp \left  \{
              i N S_N[x;j,k]
           \right \}  \>,
\end{equation*}
with
\begin{widetext}
\begin{equation}
   S_{N}[x;j,k]
   =
   S_{\text{class}}[x]
   +
   \sum_\alpha \int_{\C} \rd t \, x_\alpha(t) \, j_\alpha(t)
   +
   \frac{1}{2}
   \sum_{\alpha,\beta} \int_{\C} \rd t \!\! \int_{\C} \rd t' \,
      x_\alpha(t) \, k_{\alpha,\beta}(t,t') \, x_\beta(t')
   \>,
   \label{eq:SNxjk}
\end{equation}
where
\begin{align}
   S_{\text{class}}[x]
   &=
   - \frac{1}{2} \sum_{\alpha,\beta}
     \int_{\C} \rd t \!\! \int_{\C} \rd t' \,
   x_\alpha(t) \, \Delta_{\alpha,\beta}^{-1}[x](t,t') \, x_\beta(t')
   =
   S_{0} + S_{\text{int}}[x]  \>,
   \label{eq:Sclass} \\
   S_{0}
   &=
   - \frac{1}{2} \sum_{\alpha,\beta}
     \int_{\C} \rd t \!\! \int_{\C} \rd t' \,
   x_\alpha(t) \, \Delta_{0\,\alpha,\beta}^{-1}(t,t') \, x_\beta(t')
   \>,
   \label{eq:S0} \\
   S_{\text{int}}[x]
   &=
   - \frac{1}{2}
     \int_{\C} \rd t \, \chi(t) \sum_{i} x_i^2(t)  \>,
   \label{eq:Sint}
\end{align}
\end{widetext}
and where $\Delta_{0\,\alpha,\beta}^{-1}(t,t')$ is given by:
\begin{align*}
   \Delta_{0\,\alpha,\beta}^{-1}(t,t')
   &=
   \begin{pmatrix}
      D^{-1}(t,t')   & 0                  \\
      0              & G_{0\,ij}^{-1}(t,t')
   \end{pmatrix}  \>,
   \\
   G_{0\,ij}^{-1}(t,t')
   &=
   \biggl \{
      \frac{\rd^2}{\rd t^2}
      \biggr \} \, \delta_{ij} \delta_{\C}(t,t') \>.
\end{align*}
with $D^{-1}(t,t') $ given by Eq.~\eqref{SD.eq:dginv}.  In this
formalism, we have separated out an ``interaction'' term,
Eq.~\eqref{eq:Sint}, which depends on the coordinates $x_\alpha(t)$,
from a bare Green function $G_{0\,ij}^{-1}(t,t')$, \emph{which is
independent of the coordinates $x_\alpha(t)$}, in contrast to our
previous definitions in Eq.~\eqref{SD.eq:dginv}.  The term $r_0^2
\chi(t)/2$ has been absorbed into the definition of the current
$\tilde{J}(t)$ in Eq.~\eqref{eq:xjextended}.

The second Legendre transform of $W[j,k]$ is the CJT effective action:
\begin{multline*}
   \Gamma[x, \G]
   =
   W[j,k] -
   \sum_\alpha \int_{\C} \rd t \, x_\alpha(t) \, j_\alpha(t)
   \\
   +
   \frac{1}{2} \sum_{\alpha,\beta}
   \int_{\C} \rd t \!\! \int_{\C} \rd t' \,
      k_{\alpha,\beta}(t,t') 
      \left  \{
         x_\alpha(t) \, x_\beta(t') + \G_{\alpha,\beta}(t,t')
      \right \}
\end{multline*}
CJT showed that $\Gamma[x, \G]$ can be obtained as a series expansion
in terms of 2-PI graphs.  That is, introducing the functional operator,
\begin{align}
   G^{-1}_{\alpha,\beta}[x](t,t')
   &=
   - \frac{\delta^2 S_{0}[x]}
          {\delta x_{\alpha}(t) \, \delta x_{\beta}(t')}
   \notag \\
   &=
   \begin{pmatrix}
      D^{-1}(t,t')   & \bar{K}_j^{-1}[x](t,t') \\
      K_i^{-1}[x](t,t') & G_{i,j}^{-1}[x](t,t')
   \end{pmatrix}  \>,
   \label{eq:scd}
\end{align}
which is the same as the $G^{-1}_{\alpha,\beta}[x](t,t')$ as defined
in Eq.~\eqref{SD.eq:ginvdef}, one can write the effective action in
the form:
\begin{multline}
   \Gamma[x,\G]
   =
   S_{\text{class}}[x] +
   \frac{i}{2} \mathrm{Tr} \{ \, \ln \, [ \, \G^{-1} \, ] \}
   \\
   +
   \frac{i}{2} \mathrm{Tr} \{ G^{-1}[x] \, \G - 1 \}
   +
   \Gamma_2[x,\G]  \>.
   \label{eq:CJT}
\end{multline}

The quantity $\Gamma_2[x,\mathcal{G}]$ has a simple graphical
interpretation in terms of all the 2-PI vacuum graphs using vertices
from the interaction term.  The Hartree and leading order in large-$N$
approximation for the $x^4$ potential was obtained by CJT using a
single two-loop vacuum graph in the $O(N)$ theory written in terms of
only the coordinates $x_i$.  Our strategy for obtaining a resummation
of the large-$N$ approximation is to first rewrite the theory in terms
of the composite field $\chi$, and the equivalent Lagrangian given in
Eq.~\eqref{SD.eq:classLiii}.  Using these new variables, we then
choose for $\Gamma_2[x,\mathcal{G}]$ the 2-PI graphs shown in
Fig.~\ref{fig:burger}, which is now written in terms of the full
$\chi$ and $x$ propagators and the trilinear coupling $\chi(t) \,
x_i^2(t)/2$.

%
%
\section{Bare Vertex Approximation}
\label{sec:BVA}

The bare vertex approximation (BVA) is obtained by setting the
vertex function equal to its bare value in the exact equations
for the one and two point functions.  This is an energy conserving
approximation which leads to integral equations for the
three-$\chi$ vertex function as well as for the $x$-$x$-$\chi$
vertex function. The bare vertex approximation consists of making
the replacement
\begin{equation}
   \Gamma_{\alpha,\beta,\gamma}[x](t_1,t_2,t_3)
   =
   f_{\alpha,\beta,\gamma} \,
   \delta_{\C}(t_1,t_2) \, \delta_{\C}(t_1,t_3) \>.
   \label{SD.eq:vertexapprox}
\end{equation}
in the exact Schwinger-Dyson equations for the self-energies,
Eqs.~\eqref{eq:allSigmas}.  This gives for the BVA:
\begin{align}
   \Pi(t,t')
   &=
   \frac{i}{2N} \sum_{ij}
   \mathcal{G}_{ij}(t,t') \, \mathcal{G}_{ji}(t',t) \>,
   \label{eq:SigmasBVA} \\
   \Omega_i(t,t')
   &=
   \frac{i}{N} \,
   \sum_j \bK_j(t,t') \, \mathcal{G}_{ji}(t,t') \>,
   \notag \\
   \bar\Omega_i(t,t')
   &=
   \frac{i}{N} \,
   \sum_j \mathcal{K}_j(t',t) \, \mathcal{G}_{ji}(t',t) \>,
   \notag \\
   \Sigma_{ij}(t,t')
   &=
   \frac{i}{N} \,
   \bigl \{
      \bK_{i}(t,t') \, \mathcal{K}_{j}(t',t)
      +
      \mathcal{G}_{ij}(t,t') \, \mathcal{D}(t',t)
   \bigr \} \>,
   \notag
\end{align}
where we have used the symmetry property, $\mathcal{G}_{ij}(t,t') =
\mathcal{G}_{ji}(t',t)$ and $\mathcal{K}_i(t,t') =
\bK_i(t',t)$.  Thus we find $\bar\Omega_i(t,t') =
\Omega_i(t',t)$.  The self-energies \eqref{eq:SigmasBVA} are then
used in Eqs.~\eqref{eq:GGinvGinvSigma} to find the one- and two-point
functions.  For the Green functions, we find:
\begin{equation}
   \mathcal{G}_{\alpha,\beta}^{-1}(t,t')
   =
   G_{\alpha,\beta}^{-1}(t,t')
   +
   \Sigma_{\text{BVA}\, \alpha,\beta}(t,t')  \>,
   \label{eq:GGinvGinvSigmaBVA}
\end{equation}
with $\Sigma_{\text{BVA}\, \alpha,\beta}(t,t')$ given by
Eq.~\eqref{eq:SigmasBVA}.  The inversion of
Eq.~\eqref{eq:GGinvGinvSigmaBVA} is given by
Eq.~\eqref{SD.eq:Gmateq}, which is a set of four coupled integral
equations for the four BVA Green functions, which must be
solved simultaneously.

From Eqs.~\eqref{SD.eq:Gammachieq} and \eqref{SD.eq:Gammaxeq}, the
equations of motion for $x_i(t)$ and the gap equation for $\chi(t)$ is
then given by:
\begin{gather}
   \biggl \{
      \frac{\rd^2}{\rd t^2}
      + \chi(t)
   \biggr \} \, x_i(t)
   +
   \frac{1}{N} \mathcal{K}_i(t,t) / i
   = 0 \>,
   \label{eq:GammaxeqBVA} \\
   \chi(t)
   =
   \frac{g}{2}
   \biggl \{
      \sum_i
      \biggl [
         x_i^2(t)
         +
         \frac{1}{N} \mathcal{G}_{ii}(t,t)/i
      \biggr ]
      -
      r_0^2
   \biggr \} \>.
   \label{eq:GammachieqBVA}
\end{gather}
For the quantum roll, we further set $x_i(t) = 0$.  This means that
$\mathcal{K}_i(t,t) = \bK_i(t,t) = 0$, so that $G_{\alpha
\beta}(t,t')$ is diagonal, and results in the following set of
equations for the Green functions:
\begin{align}
   \mathcal{D}(t,t')
   &=
   D(t,t')
   \notag \\ 
   &\!\!\!\!\!\!\!\!\!\!\!\!-
   \int_{\C} \rd t_1 \!\! \int_{\C} \rd t_2 \,
   D(t,t_1) \, \Pi(t_1,t_2) \, \mathcal{D}(t_2,t') \>,
   \label{SD.eq:DDintzerox}
   \\
   \mathcal{G}_{ij}(t,t')
   &=
   G_{ij}(t,t')
   \notag \\
   &\!\!\!\!\!\!\!\!\!\!\!\!-
   \sum_{kl} \int_{\C} \rd t_1 \!\! \int_{\C} \rd t_2 \,
   G_{ik}(t,t_1) \, \Sigma_{kl}(t_1,t_2) \, \mathcal{G}_{lj}(t_2,t') \>,
   \label{SD.eq:GGinyzerox}
\end{align}
where
\begin{align}
   \Pi(t,t')
   &=
   \frac{i}{2N} \sum_{ij}
   \mathcal{G}_{ij}(t,t') \, \mathcal{G}_{ji}(t',t) \>,
   \notag \\
   \Sigma_{ij}(t,t')
   &=
   \frac{i}{N} \, \mathcal{G}_{ij}(t,t') \, \mathcal{D}(t',t) \>.
   \label{SD.eq:selfxqero}
\end{align}
The gap equation for $\chi(t)$ becomes:
\begin{equation}
   \chi(t)
   =
   \frac{g}{2}
   \biggl \{
      \frac{1}{N} \sum_i \mathcal{G}_{ii}(t,t)/i
      -
      r_0^2
   \biggr \}  \>.
   \label{SD.eq:chixzero}
\end{equation}
In addition, for this case, the initial conditions imply that we can
take $G_{ij}(t,t')$ and $\mathcal{G}_{ij}(t,t')$ to be diagonal, which
greatly simplify the integral equations.  The BVA for the quantum roll
requires that we solve equations \eqref{SD.eq:DDintzerox},
\eqref{SD.eq:GGinyzerox}, \eqref{SD.eq:selfxqero}, and
\eqref{SD.eq:chixzero} simultaneously using the numerical methods
described in refs.~\cite{ref:MDC} and \cite{ref:BMIM}.

Because of the interest in using the BVA approximation in QED (and
QCD) plasma conductivity problems, we will discuss the integral
equation one obtains for the vertex function in what follows.  It
was precisely because this approximation gives the sum of the
graphs used in non-relativistic plasmas (see Fig.~\ref{fig:BVA})
in conductivity calculations which gave both accurate results as
well as giving physical answers that initially interested us in
this approximation.

The three-point vertex functions for the BVA are given by functional
differentiation of the inverse two point functions:
\begin{align}
   \Lambda(t_1,t_2,t_3)
   &\equiv
   \Gamma_{000}(t_1,t_2,t_3)
   =
   \frac{\delta \, \mathcal{D}^{-1}(t_1,t_2)}
        {\delta \chi(t_3)}
   \label{SD.eq:Gamma000threezero} \\
   \Gamma_{ij}(t_1,t_2,t_3)
   &\equiv
   \Gamma_{ij0}(t_1,t_2,t_3)
   =
   \frac{\delta \, \mathcal{G}_{ij}^{-1}(t_1,t_2)}
        {\delta \chi(t_3)}  \>,
   \label{SD.eq:Gammaij0threezero}
\end{align}
and obtain the coupled integral equations:
\begin{widetext}
\begin{equation}
   \Lambda(t_1,t_2,t_3)
   =
   -
   \frac{i}{N}
   \int_{\C} \rd t_4 \int_{\C} \rd t_5 \sum_{ijkl}
      \mathcal{G}_{ik}(t_1,t_4) \,
      \Gamma_{kl}(t_4,t_5,t_3) \,
      \mathcal{G}_{lj}(t_5,t_2) \,
      \mathcal{G}_{ji}(t_2,t_1)
   \label{SD.eq:verteqa}
\end{equation}
and
\begin{multline}
   \Gamma_{ij}(t_1,t_2,t_3)
   =
   \delta_{ij} \, \delta_{\C}(t_1,t_2) \, \delta_{\C}(t_1,t_3)
   \\
   -
   \int_{\C} \rd t_4 \int_{\C} \rd t_5
   \biggl \{
      \sum_{kl}
      \mathcal{G}_{ik}(t_1,t_4) \,
      \Gamma_{kl}(t_4,t_5,t_3) \,
      \mathcal{G}_{lj}(t_5,t_2) \,
      \mathcal{D}(t_2,t_1)
   +
      \mathcal{G}_{ij}(t_1,t_2)
      \mathcal{D}(t_2,t_4) \,
      \Lambda(t_4,t_5,t_3) \,
      \mathcal{D}(t_5,t_1)
   \biggr \} \>.
   \label{SD.eq:verteqb}
\end{multline}

This is shown diagrammatically in Fig.~\ref{fig:BVA}.  Looking at
the diagrams, if we iterate these equations, we sum all the
``rainbow'' diagrams.  As advertised, comparing these graphs with
those shown in Fig.~\ref{fig:SDV}, $\K_1$ is approximated in the
BVA by $\chi$ exchange and $\K_2$ by $x$ exchange in the
$t$-channel.
%
%
\begin{figure}[t]
   \centering
   \includegraphics[width=3.0in]{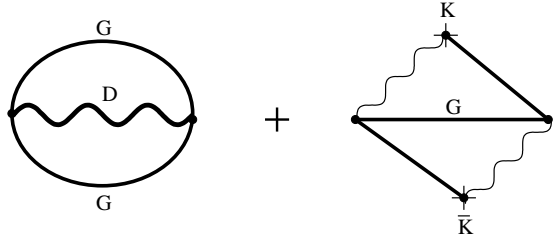}
   \caption{Vacuum graphs contributing to the 2PI part of the effective
    action
   $\Gamma_2[\mathcal{G}]$.  Solid lines represent the
   $\mathcal{G}_{ij}(t,t')$ propagator, the wiggly to solid
   lines represent the
   $\mathcal{K}_i(t,t')$ and $\bK_i(t,t')$
   propagator, and wiggly lines are the
   $\mathcal{D}(t,t')$ propagator.}
   \label{fig:burger}
\end{figure}
%
%

Let us show that this approximation is easy to obtain from the CJT
formalism once we treat $\G$ and $\D$ and $\K$ on exactly the same
footing.  We choose for our approximation to $\Gamma_2[\G]$ the 2-PI
graphs shown in Fig.~\ref{fig:burger}.  This gives:
\begin{equation}
   \Gamma_2[\G]
   =
   - \frac{1}{4 N} \sum_{ij} \int_\C \rd t_1 \int_\C \rd t_2 \,
     \D(t_1,t_2) \, \G_{ij}(t_1, t_2) \, \G_{ji}(t_2,t_1)
   - \frac{1}{2 N} \sum_{ij}
      \int_\C \rd t_1 \int_\C \rd t_2 \,
      \bK_i(t_1,t_2) \, \G_{ij}(t_1,t_2) \,
      \mathcal{K}_j(t_2,t_1)  \>.
   \label{eq:Gamma2CJT}
\end{equation}
\end{widetext}
Since the $\D$ propagator sums the contact term plus all the
polarization bubbles $\Pi$ of the original quartic interaction $g
x^4$, if we reexpand $\D$ in a power series in $\Pi$ then the first
two terms in the series give the graphs used in the approximation
of\cite{ref:Hu} and\cite{ref:berges}.  The CJT action is given by
Eq.~\eqref{eq:CJT}.  The stationary condition for
$\G_{\alpha,\beta}(t,t')$ gives:
\begin{equation*}
   \frac{\delta \Gamma[x,\G]}{\delta \G_{\alpha \beta}}
   =
   \frac{i}{2} \left \{ G^{-1}_{\alpha \beta} [x]
   -
   \G^{-1}_{\alpha\beta} \right \}
   +
   \frac{\delta \Gamma_2[\G]}{\delta \G_{\alpha \beta}}
   = 0 \>,
\end{equation*}
or
\begin{equation*}
   \mathcal{G}_{\alpha,\beta}^{-1}(t,t')
   =
   G_{\alpha,\beta}^{-1}(t,t')
   +
   \Sigma_{\text{BVA}\,
   \alpha,\beta}[\G](t,t') \>,
\end{equation*}
where:
\begin{equation}
   \Sigma_{\text{BVA}\, \alpha,\beta}[\G](t,t')
   =
   -2i \, \frac{\delta \Gamma_2[\G]}
               {\delta \G_{\alpha \beta}(t,t')} \>.
   \label{eq:GinvCJTBVA}
\end{equation}
Carrying out the derivatives of $\Gamma_2[\G]$ given in
Eq.~\eqref{eq:Gamma2CJT}, we find that $\Sigma_{\text{BVA}\,
\alpha,\beta}(t,t')$ is exactly the same as found in
Eq.~\eqref{eq:SigmasBVA} using the Schwinger-Dyson equations in the
BVA approximation.  The stationary condition for $x_\alpha$ also gives
the same equations of motion for $x_i(t)$ and gap equation for
$\chi(t)$ as found in Eqs.~\eqref{eq:GammaxeqBVA} and
\eqref{eq:GammachieqBVA} using the Schwinger-Dyson equations in the
BVA.  Thus we conclude that the CJT action, as given in
Eqs.~\eqref{eq:CJT} and \eqref{eq:Gamma2CJT}, gives exactly the same
set of equations as in the Schwinger-Dyson BVA truncation.

The energy for the BVA is obtained from \eqref{eq:fullenergy} by
using \eqref{SD.eq:vertexapprox} for the vertex function.  We find:
\begin{widetext}
\begin{align}
   E/N
   &=
   \frac{1}{2}
      \sum_i
      \biggl \{
         \left  . \frac{ \partial^2 \, \G(t,t')/i }
                  { \partial t \, \partial t' }
         \right |_{t'=t}
         +
         \chi(t) \, \G(t,t)/i
         -
         r_0^2 \chi(t)
         -
         \frac{1}{g}
         \left  [ 
            \chi^2(t) + \frac{1}{N}\D(t,t)/i 
         \right ]
    \notag \\
    & \quad
    +
    \frac{1}{N} \sum_{ij}
    \int_{\C} \rd t_1 \,
    \D(t_1,t) \G_{ij}(t_1,t) \G_{ji}(t,t_1)
      \biggr \}  \>.
    \label{eq:BVAenergy}
\end{align}
\end{widetext}
where, for our case, we have set $x_i(t) = \dot{x}_i(t) = 0$.  Since
the BVA equations are derived from an effective action, energy is
conserved.

%
%
\section{Dynamical Debye Screening Approximation}
\label{sec:DDSA}

In plasma studies of the electric conductivity of fully ionized
plasmas \cite{ref:plasma1,ref:plasma2}, it was found that in
order to correctly determine the conductivity it was necessary to
have an approximation where the photon propagator included the
effects of dynamical Debye screening in the random phase
approximation.  This improved propagator was then used in a
scattering kernel in the kinetic equations. In our model, the
$\chi$ field plays the roll of the photon in the dynamics of the
$x_i$ oscillators.  The lowest approximation that includes the
polarization effects in $\D$ is precisely the leading order in
large-$N$ approximation to $\D$, namely $\D_0$ (see Eq.
\ref{eq:dloln}) which is discussed below in our derivation of the
NLOLN approximation . The leading order in large-$N$
approximation is similar in spirit to the random phase
approximation.  The equation for $\D^{-1}(t,t')$ in leading order
in large-$N$ is given by:
\begin{equation}
   \D_0^{-1}(t,t')
   =
   -\frac{1}{g} \delta_{\C}(t,t') + \Pi_0(t,t') \>,
   \label{eq:DN}
\end{equation}
where
\begin{multline*}
   \Pi_0(t,t')
   =
   \frac{i}{2N} \, \sum_{i,j} G_{ij}(t,t') \, G_{ji}(t',t)
   \\
   + \sum_{i,j} \, x_i(t) \, G_{ij}(t,t') \, x_j(t') \>.
\end{multline*}

In the QED plasma problem, the $\chi$ propagator becomes the photon
propagator and the delta function in $\D_0$ is replaced by the
bare photon propagator.  It is the bubble in $\Pi_0$ that leads
to the Debye screening of the photon.  It is because of our
interest in QED that we call this approximation the DDSA.

Let us now specialize to the case when $x_i(t) = 0$.  The equation for
the full $x$ propagator ${\cal G}$ is:
\begin{multline}
   \G_{ij} (t,t')
   =
   G_{ij}(t,t')
   \\
   -
   \sum_{k,l}
      \int_{\C} \! {\rm d}t_1 \!\! \int_{\C} \! {\rm d}t_2 \,
      G_{ik}(t,t_1) \, \Sigma_{kl}(t_1,t_2) \, \G_{lj}(t_2,t') \>,
   \label{eq:Gfulli}
\end{multline}
with the self energy depending on the full ${\cal G}$ and the leading
order in $1/N$ approximation to $\mathcal{D}$ given by
Eq.~\eqref{eq:DN}:
\begin{equation}
   \Sigma_{kl}(t,t')
   =
   \frac{i}{N} \, \G_{kl}(t,t') \, D(t,t')
   \>.
   \label{eq:SigmaDDSA}
\end{equation}
The gap equation is:
\begin{equation}
   \chi(t)
   =
   \frac{g}{2}\
   \left  \{
      \sum_i \frac{1}{N} \mathcal{G}_{ii}(t,t)/i
       - r_0^2
   \right \}
   \>.
   \label{eq:chiagain}
\end{equation}
There is a nontrivial vertex function in this approximation given by:
\begin{widetext}
\begin{align}
   \Gamma_{ij}(t_1,t_2,t_3)
   &=
   \frac{\delta \G^{-1}_{ij}[\chi](t_1,t_2)}{\delta \chi(t_3)}
   \notag \\
   &=
   \delta_{\C}(t_1,t_2) \delta_{\C}(t_3,t_2) \delta_{ij}
   -
   \sum_{kl} \int_\C \rd t_4 \int_\C \rd t_5 \,
   \Gamma_{kl}(t_4,t_5,t_3) \,
   \G_{ki}(t_4,t_1) \, D(t_1,t_2) \, \G_{jl}(t_2,t_5)
   \notag \\
   & \qquad
   -
   \int_\C \rd t_4 \int_\C \rd t_5 \,
   \Lambda(t_4,t_5,t_3) \,
   D(t_4,t_1) \, \G_{ij}(t_1,t_2) \, D(t_2,t_5) \>.
\end{align}
\end{widetext}
This equation can be obtained from the exact integral equation for
$\Gamma$ shown pictorially in Fig.~\ref{fig:SDV} by making two
approximations.  The first is to approximate the exact
three-$\chi$ vertex function by the triangle graph, which is the
leading term in the $1/N$ expansion of this function.  The second
is to replace the scattering kernels, $K_1$ and $K_2$ by single
particle exchange in the t-channel. The reason for our studying
this approximation is that, the same approximation made in QED
can be shown to be the lowest order resummation scheme that
preserves Ward identities (\cite{ref:emil}).

The DDSA approximation can be derived from an effective action by
modifying slightly the approach of Cornwall, Jackiw and Tomboulis
(CJT)\cite{ref:CJT}. The discussion that follows here is due to Emil
Mottola and Luis Bettencourt\cite{ref:emil}.  Thinking of the fields
$x$ and $\chi$ as part of an $N+1$ component field, and considering
the case that $\langle \hat{x}(t) \rangle = 0$ where there is no mixed
propagator, one can write a CJT like action for the generating
functional of the twice Legendre transformed effective action as:
\begin{multline}
   \Gamma[\chi,\G,\D]
   =
   S_{\text{class}}[\chi]
   +
   \frac{i}{2} \mathrm{Tr} \{ \, \ln \, [ \, \D^{-1} \, ] \}
   \\
   +
   \frac{i}{2} \mathrm{Tr} \{ \, \ln \, [ \, \G^{-1} \, ] \}
   +
   \frac{i}{2} \mathrm{Tr}
      \{ \D_0^{-1} \, \D + G^{-1}[\chi] \, \G - 1 \}
   +
   \Gamma_2[\G] \>.
   \label{eq:GammaDDSA}
\end{multline}
here $G^{-1}(t,t')$ is defined by \eqref{SD.eq:dginv} and $\D_0(t,t')$
by Eq.~\eqref{eq:DN}.  $\D_0(t,t')$ is considered an \emph{external}
parameter, and is not varied to obtain the equations of motion.
In the DDSA, the 2-PI contribution to the action,
$\Gamma_2[\G]$, for the case when $x_i(t)=0$, is given by
Eq.~\eqref{eq:Gamma2CJT} with $\D(t,t')$ set equal to its LOLN
value $\D_0(t,t')$:
\begin{multline}
   \Gamma_2[\G]
   =
   \\
   - \frac{1}{4 N}
   \sum_{ij} \int_\C \rd t_1 \int_\C \rd t_2 \,
      \D_0(t_1,t_2) \, \G_{ij}(t_1,t_2) \, \G_{ji}(t_2,t_1)  \>.
\end{multline}

By varying the action (\ref{eq:GammaDDSA}), we reproduce
Eqs.~\eqref{eq:Gfulli} and \eqref{eq:chiagain}.  Although there
is an effective action for the DDSA approximation, since $\D_0$
is treated as an external time-dependent propagator, energy
conservation is violated at order $1/N$.  At modest $N$ we will
find that this causes this approximation to become inaccurate
after several oscillations. However, it is more accurate at these
modest values of $N$ than the LOLN approximation, as well
avoiding the unboundedness of the NLOLN approximation we discuss
next.

%
%
\section{The large-$N$ approximation}
\label{sec:largeN}

The large-$N$ expansion is obtained from Eq.~\eqref{SD.eq:Z} by first
integrating over all the $x_i$ and then evaluating the remaining
functional integral over $\chi$ by steepest descent.  The effective
action, as a power series in $1/N$, is obtained from the first
Legendre transform of the generating functional.  In a previous
paper\cite{ref:ctpN}, we obtained equations for the next to leading
order large-$N$ approximation (NLOLN) to the action, and gave
numerical results for the quantum roll.  For completeness, we review
those equations here.  To order $1/N$, we obtain:
\begin{multline*}
   \Gamma_{\text{Large-N}}[x]
   =
   S_{\text{class}}[x]
   \\
   +
   \int_{\C} \rd t \,
       \biggl \{
          \frac{i}{2} \sum_i \,
          \ln \, [ G_{ii}^{-1}(t,t) ]
          + \frac{i}{2 N} \ln [ \mathcal{D}^{-1}_{0}(t,t) ]
       \biggr\}
   \>,
\end{multline*}
where $S_{\text{class}}[x]$ is given by
Eq.~\eqref{eq:Sclass}, and $\mathcal{D}^{-1}_{0}(t,t')$
is the inverse propagators for $\chi$ in lowest order in
the $1/N$ expansion, given by
\begin{equation}
   \mathcal{D}^{-1}_{0}(t,t')
   =
      D^{-1}(t,t') + \Pi_0(t,t') \>, \label{eq:dloln}
\end{equation}
with
\begin{align}
   \Pi_0(t,t')
   &=
   \frac{i}{2 N} \, \sum_{i,j} G_{ij}(t,t') \, G_{ji}(t',t)
   \notag \\
   &
   - \sum_{i,j} \, x_i(t) \, G_{ij}(t,t') \, x_j(t') \>.
   \label{eq:PilrgNdef}
\end{align}
Here $D^{-1}(t,t')$ and $G^{-1}_{ij}(t,t')$ are the same as
Eqs.~\eqref{SD.eq:dginv} that we defined earlier.

The equations of motion for the classical fields $x_i(t)$, to next to
leading order in $1/N$, are given by:
\begin{multline}
   \left  \{
      \frac{{\rm d}^2 }
           {{\rm d} t^2 }
      + \chi(t)
   \right\} x_i(t)
   \\
   + i \, \sum_{j} \int_{\C} {\rm d}t' \,
         G_{ij}(t,t') \, \mathcal{D}_{0}(t,t') \, x_j(t')
   = 0
   \>,
\label{eq:phieom}
\end{multline}
with the gap equation for $\chi(t)$ given by
\begin{equation}
   \chi(t)
   =
   \frac{g}{2}
   \left  \{
      \sum_i \left  (
                x_i^2(t) +
                \frac{1}{N} \sum_i \G^{(2)}_{ii}(t,t) / i
             \right )
      - r_0^2
   \right \} \>,
\label{eq:Chieqn}
\end{equation}
and where the second order $x_i$ propagator
$\mathcal{G}^{(2)}_{ij}(t,t)$ and self energy $\Sigma_{ij}(t,t')$ to
order $1/N$ is given by:
\begin{multline}
   \G^{(2)}_{ij}(t,t')
   =
   G_{ij}(t,t')
   \\ 
   - \sum_{k,l}
      \int_{\C} {\rm d}t_1 \!\! \int_{\C} {\rm d}t_2 \,
      G_{ik}(t,t_1) \, \Sigma_{kl}(t_1,t_2) \, G_{lj}(t_2,t') \>,
   \label{eq:Gfull}
\end{multline}
where
\begin{equation*}
   \Sigma_{ij}(t,t')
   =
   \frac{i}{N} \, G_{ij}(t,t') \, \mathcal{D}_{0}(t,t')
     - x_i(t) \, \mathcal{D}_{0}(t,t') \, x_j(t')
   \>.
\end{equation*}
We see here that the equation for $\mathcal{G}$ is the expansion of
the BVA equation in a series of $1/N$, truncated at first order.

Let us now specialize to the case of the quantum roll problem where
$x_i(t) =0$.  In that case the two point inverse propagator for the $x$
field is
\begin{align*}
   \G^{-1}_{ij}[\chi](t_1,t_2)
   &=
   \frac{\delta^2 \Gamma_{\text{Large-N}}[x,\chi]}
        {\delta x_i(t_1) \, \delta x_j(t_2)}
   \\
   &=
   G^{-1}_{ij}[\chi](t_1,t_2) + \Sigma_{ij}[\chi](t_1,t_2) \>,
\end{align*}
with
\begin{equation*}
   \Sigma_{ij}[\chi](t,t')
   =
   \frac{i}{N} \, G_{ij}(t,t') \, \mathcal{D}_{0}(t,t')
\end{equation*}
However it is $\G^{(2)}_{ij}(t,t')$ which enters into
Eq.~\eqref{eq:Chieqn} and not $\G_{ij}(t,t')$.  Thus the solution for
$\G_{ij}(t,t')$, which we might interpret as $\langle \hat{x}_i(t)
\hat{x}_j(t') \rangle$, does not enter into the dynamics of the
solution!  \emph{This} $\G_{ii}(t,t)$ is positive definite, but
quickly blows up.

The vertex function $\Gamma_{ij}(t_1,t_2,t_3)$ is given by:
\begin{widetext}
\begin{align}
   \Gamma_{ij}(t_1,t_2,t_3)
   &=
   \frac{\delta \G^{-1}_{ij}[\chi](t_1,t_2)}{\delta \chi(t_3)}
   \\
   &=
   \delta_{\C}(t_1,t_2) \delta_{\C}(t_2,t_3) \delta_{ij}
   - \frac{i}{N} \,
   G_{ij}(t_1,t_3) \, G_{ji}(t_3,t_2) \, \D_{0}(t_2,t_1)
   \notag \\
   & \qquad
   - \frac{i}{N} \,
   \int_\C \rd t_4 \!\! \int_\C \rd t_5 \,
   G_{ij}(t_1,t_2) \, \D_{0}(t_1,t_4)
   \Lambda_0(t_4,t_5,t_3) \, \D_{0}(t_5,t_2)  \>,
   \notag 
\end{align}
where the lowest order in $1/N$ 3-$\chi$ vertex is given by
\begin{align*}
   \Lambda_0(t_4,t_5,t_3)
   &=
   \frac{\delta \D^{-1}_0(t_4,t_5)}{\delta \chi(t_3)}
   \\
   &=
   - \frac{i}{N} \sum_{ijk}
   G_{ij}(t_4,t_3) \, G_{kl}(t_3,t_5) \, G_{li}(t_5,t_4)
   \>.
\end{align*}
We immediately see that this is not an integral equation but again, is
the lowest order in $1/N$ contribution to Eq.~\eqref{SD.eq:verteqa}.

The inverse $\chi$ propagator gets $1/N$ corrections which are of two
types, one is a self energy correction to the $x$ propagator and the
other is a new three loop graph containing two lowest order $\chi$
propagators.  We find
\begin{align*}
   \D^{-1}(1,2)
   &=
   \frac{\delta^2 \Gamma_{\text{Large-N}}[x,\chi]}
        {\delta \chi(t_1) \, \delta \chi(t_2)}
   \\
   &=
   - \frac{1}{g} \delta_{\C}(1,2)
   - \Pi_0(1,2)
   - \sum_{ijkl} \int_\C \rd t_3 \!\! \int_\C \rd t_4 \,
      G_{ij}(t_1,t_3) \, \Sigma_{jk}(t_3,t_4) \,
      G_{kl}(t_4,t_2) \, G_{li}(t_2,t_1)
   \\ &\qquad\qquad
   +
   \int_\C \rd t_3 \!\! \int_\C \rd t_4 \!\!
   \int_\C \rd t_5 \!\! \int_\C \rd t_6 \,
      \Lambda_0(t_4,t_1,t_3) \, \D_0(t_3,t_5) \,
      \Lambda_0(t_5,t_2,t_6) \, \D_0(t_6,t_4) \>.
\end{align*}
\end{widetext}
The last term in this equation is a $1/N$ correction to the vertex
function.  However, it is $\mathcal{D}_0$ and not $\mathcal{D}$ which
enters Eq.~\eqref{eq:Gfull}, so that the BVA and the $1/N$ expansion
will differ only by terms of order $1/N^2$.  The BVA approximation
treats $x$ and $\chi$ on exactly the same footing, whereas the
large-$N$ expansion treats $x$ exactly, but then expands in loops of
$\chi$.  So at order $1/N^2$, the large-$N$ approximation will contain
graphs omitted from the BVA approximation, and vice-versa.

%
%
\section{Results and Conclusions}
\label{sec:results}

In this section we present the results of exact numerical simulations
of the quantum roll, using initial conditions described in our
previous paper on the large-$N$ approximation\cite{ref:paper1}.  We
choose as our dimensional mass scale the second derivative of $U(r)$
at the minimum of the effective one dimensional potential $U(r)$.
This mass scale was chosen to have value $M^2=2$.  In terms of this
mass scale, the coupling constant as well as the rescaled $r_0$ are of
order one for all $N$.  The exact manner in which $g$ and $r_0$ runs
with $N$ is described in ref.~\cite{ref:paper1}.

%
%
%
\begin{figure*}[tp]
   \centering
   \includegraphics[width=5.5in]{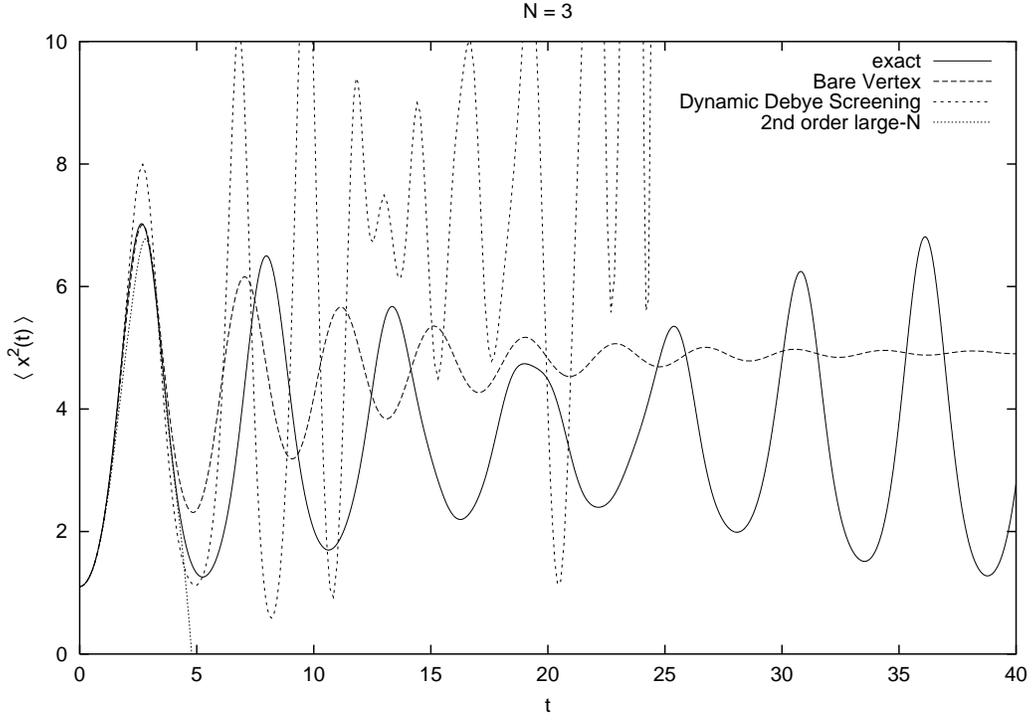}
   \caption{Plot of $\langle x^2(t) \rangle$ as a function of $t$,
   comparing the bare vertex, the dynamic Debye screening, and
   the large-$N$
   approximations to the exact solution, for $N=3$.}
   \label{SD.fig:n3_x2}
\end{figure*}
\begin{figure*}
   \centering
   \includegraphics[width=5.5in]{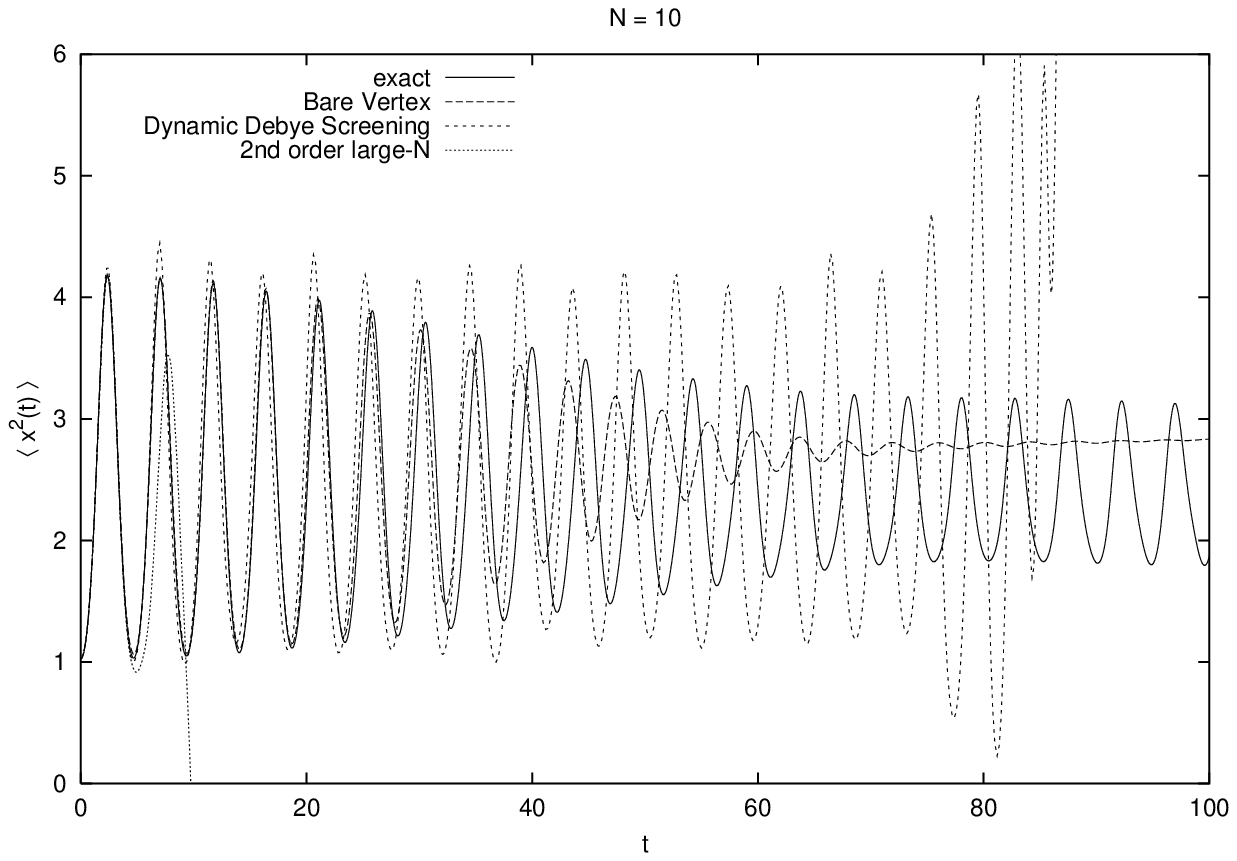}
   \caption{Plot of $\langle x^2(t) \rangle$ as a function of $t$,
   comparing the bare vertex, the dynamic Debye screening, and
   the large-$N$
   approximations to the exact solution, for $N=10$.}
   \label{SD.fig:n10_x2}
\end{figure*}
\begin{figure*}
   \centering
   \includegraphics[width=5.5in]{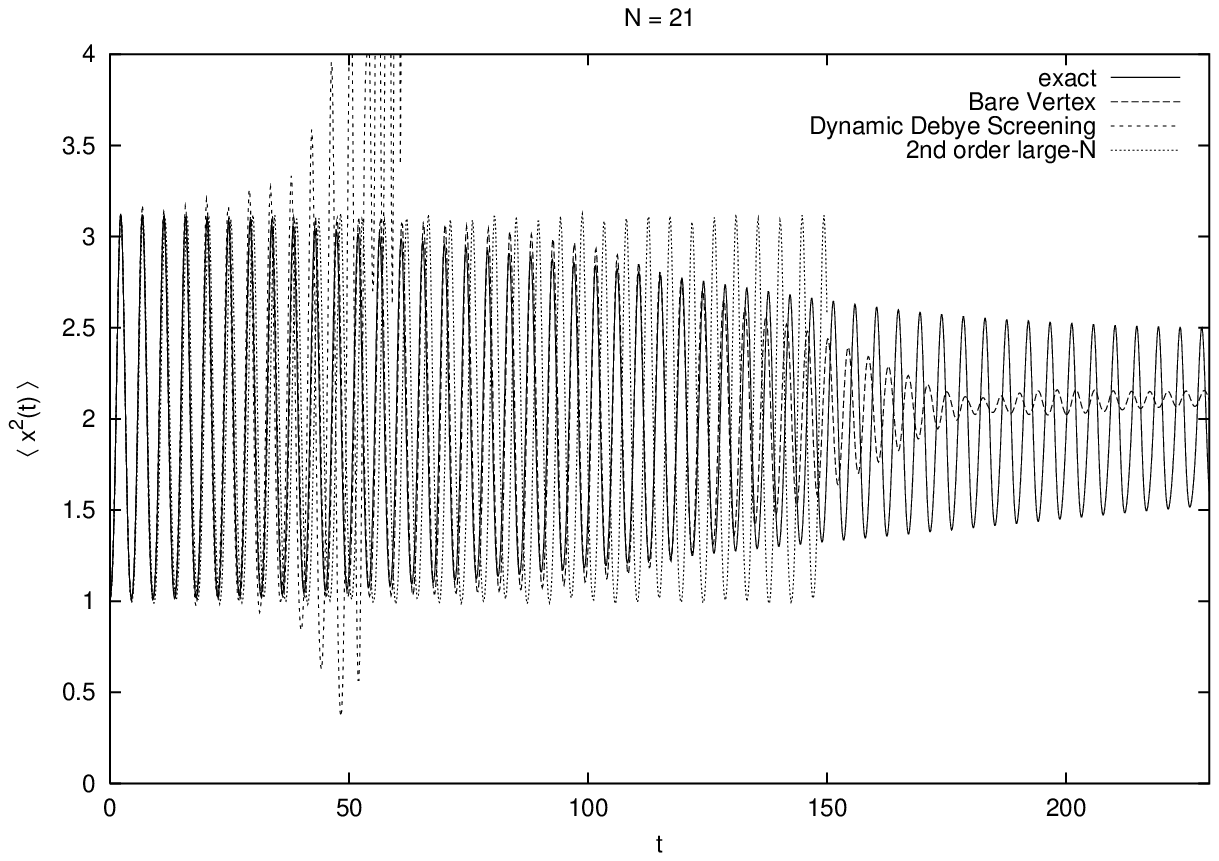}
   \caption{Plot of $\langle x^2(t) \rangle$ as a function of $t$,
   comparing the bare vertex, the dynamic Debye screening, and
   the large-$N$
   approximations to the exact solution, for $N=21$.}
   \label{SD.fig:n21_x2}
\end{figure*}
%
%
\begin{figure*}
   \centering
   \includegraphics[width=5.5in]{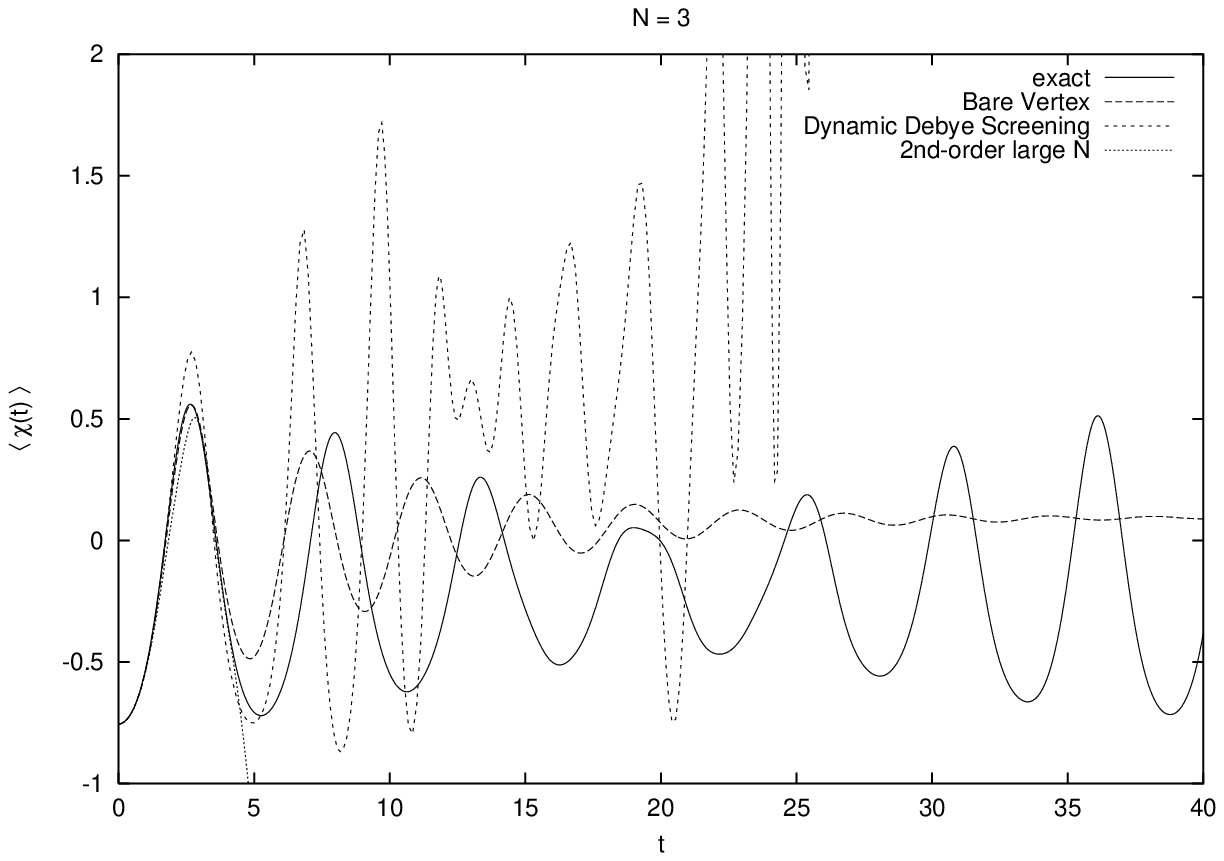}
   \caption{Plot of $\langle \chi(t) \rangle$ as a function of $t$,
   comparing the bare vertex, the dynamic Debye screening, and
   the large-$N$
   approximations to the exact solution, for $N=3$.}
   \label{SD.fig:n3_chi}
\end{figure*}
\begin{figure*}
   \centering
   \includegraphics[width=5.5in]{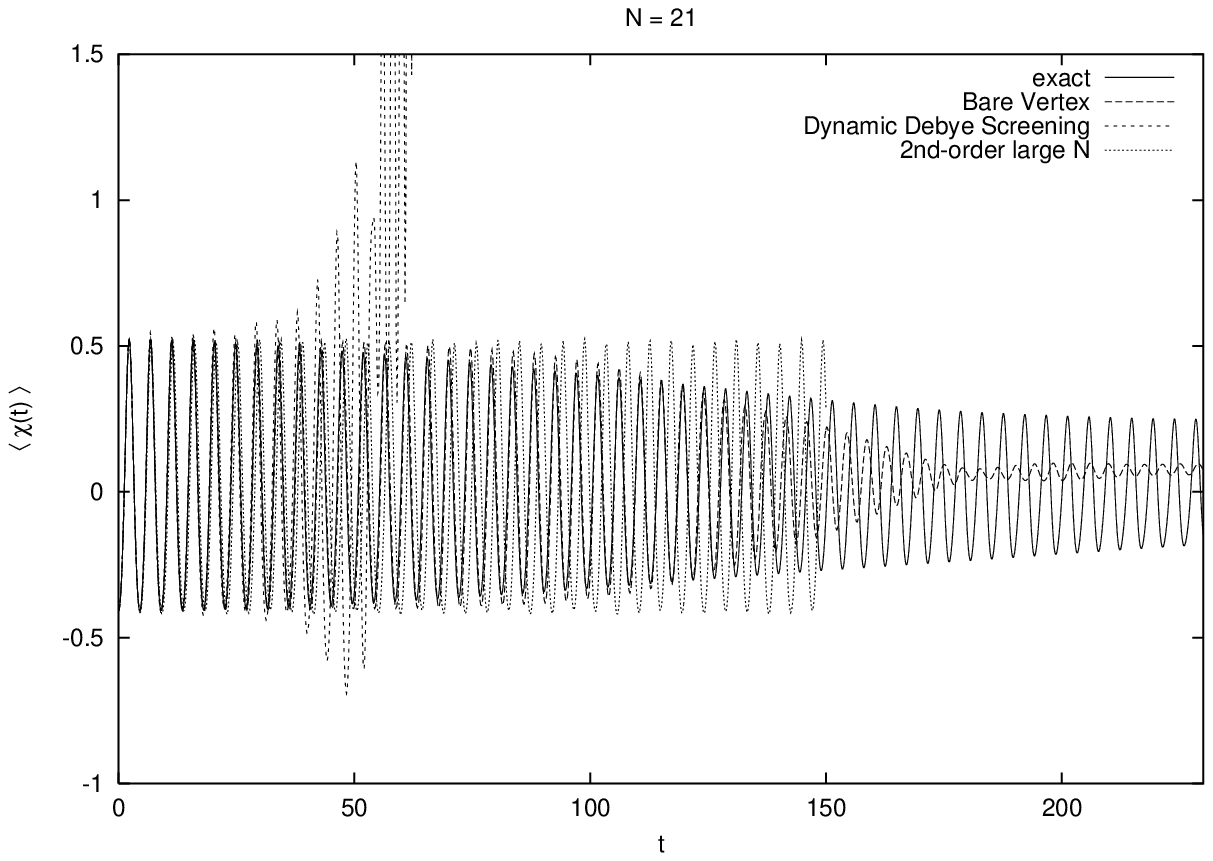}
   \caption{Plot of $\langle \chi(t) \rangle$ as a function of $t$,
   comparing the bare vertex, the dynamic Debye screening, and
   the large-$N$
   approximations to the exact solution, for $N=21$.}
   \label{SD.fig:n21_chi}
\end{figure*}
%
%
%
\begin{figure*}
   \centering
   \includegraphics[width=5.5in]{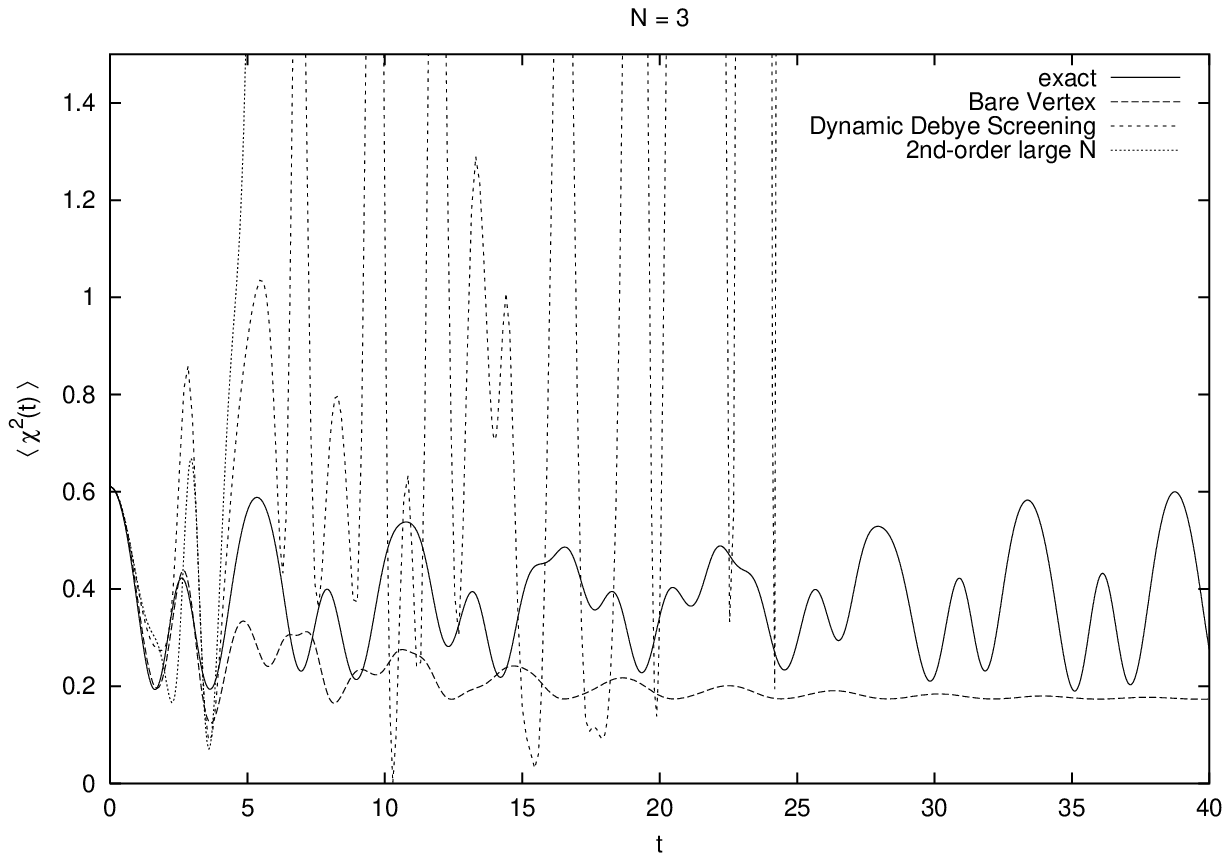}
   \caption{Plot of $\langle \chi^2(t) \rangle$ as a function of $t$,
   comparing the bare vertex, the dynamic Debye screening, and
   the large-$N$
   approximations to the exact solution for $N=3$.}
   \label{SD.fig:n3_chi2}
\end{figure*}
\begin{figure*}
   \centering
   \includegraphics[width=5.5in]{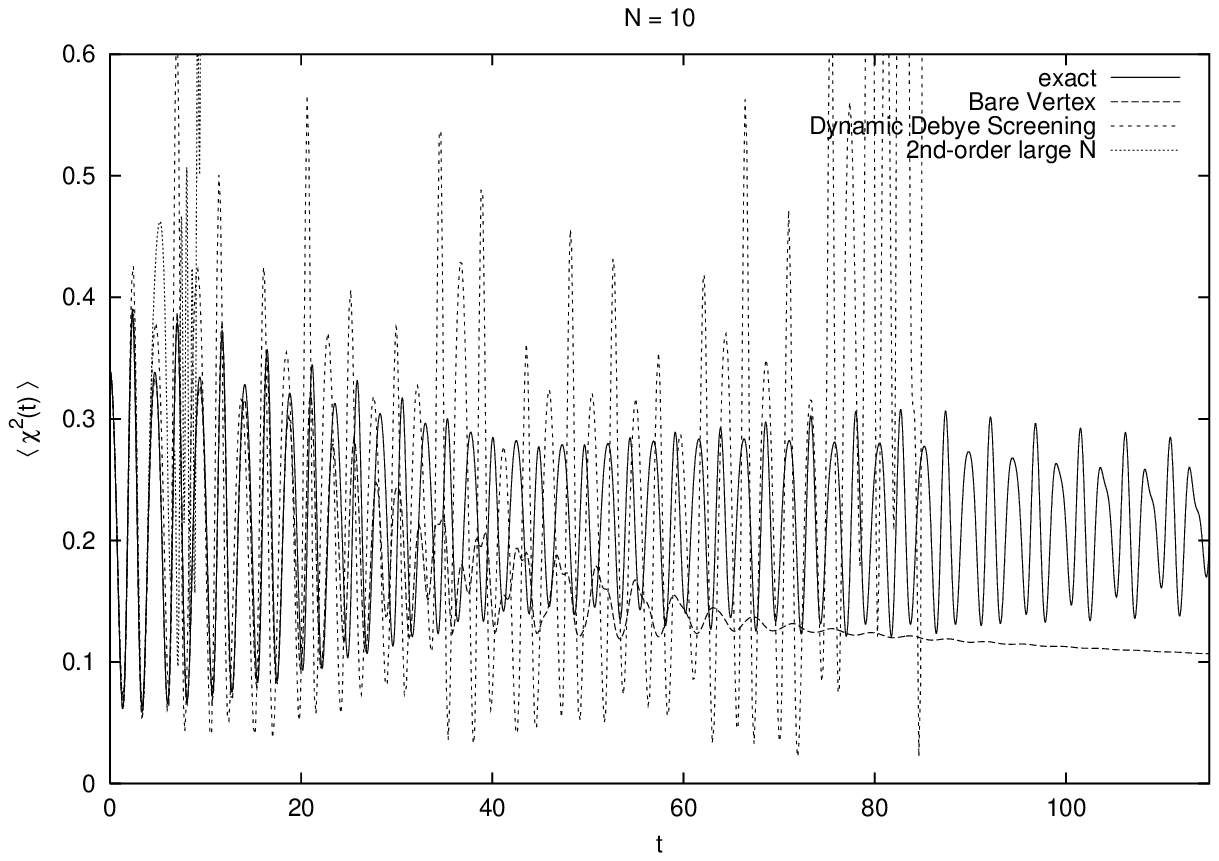}
   \caption{Plot of $\langle \chi^2(t) \rangle$ as a function of $t$,
   comparing the bare vertex, the dynamic Debye screening, and
   the large-$N$
   approximations to the exact solution for $N=10$.}
   \label{SD.fig:n10_chi2}
\end{figure*}
\begin{figure*}
   \centering
   \includegraphics[width=5.5in]{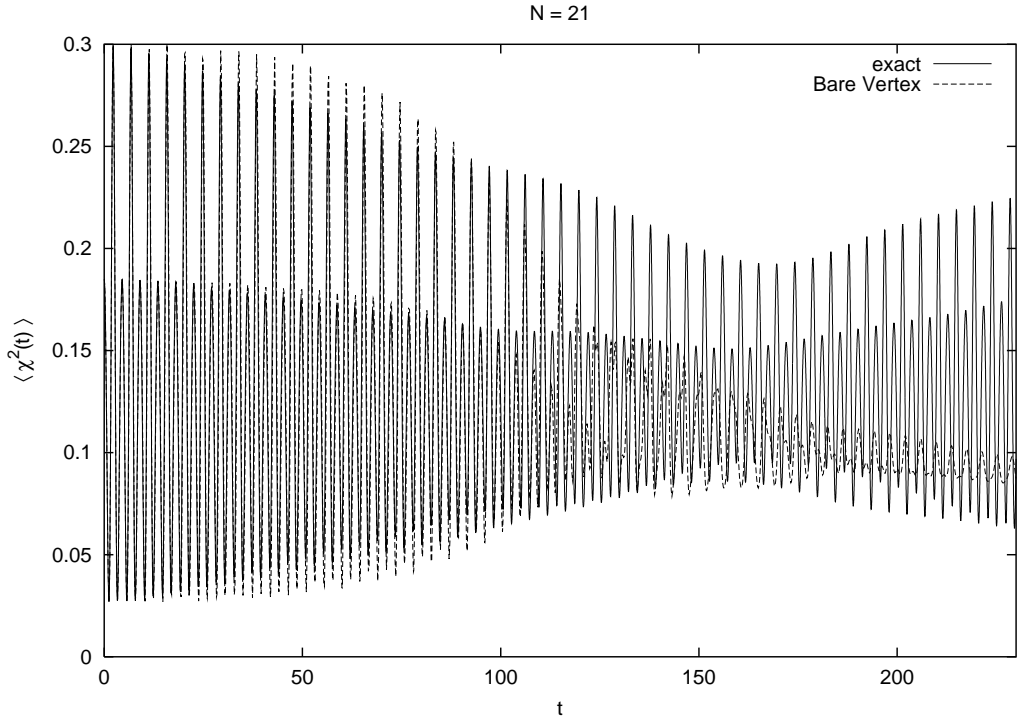}
   \caption{Plot of $\langle \chi^2(t) \rangle$ as a function of $t$,
   comparing the bare vertex
   approximation to the exact solution for $N=21$.}
   \label{SD.fig:n21_chi2}
\end{figure*}
%
%
%
\begin{figure*}
   \centering
   \includegraphics[width=5.5in]{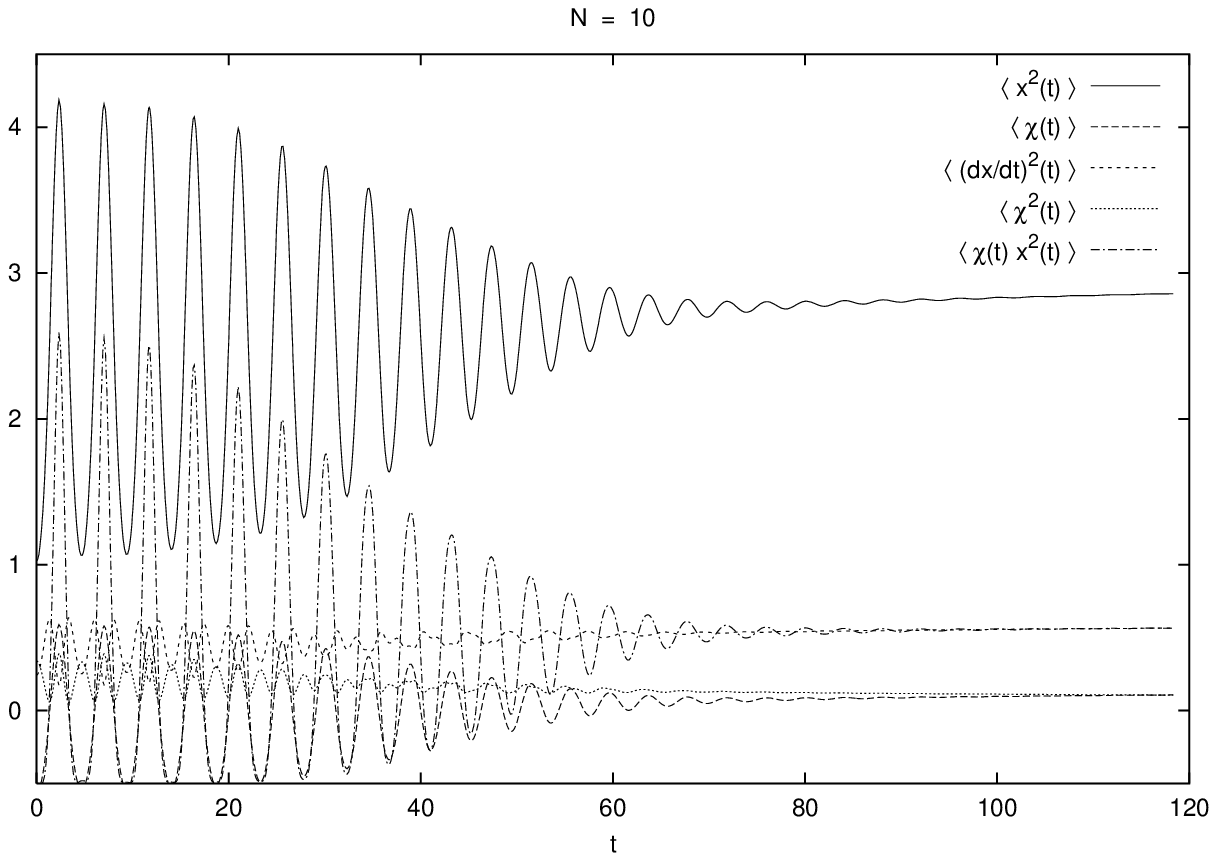}
   \caption{Plot of various contributions to the energy for the
   bare vertex approximation as a function of $t$
   for $N=10$.}
   \label{fig:energies}
\end{figure*}
%
%

As $N \rightarrow \infty$ the Hartree and leading order large-$N$
approximation become exact and an initially gaussian wave packet
remains gaussian with width equal to $\langle x^2(t) \rangle$
oscillating in a known manner.  At modest $N$, $10 < N < 20 $ an
initially Gaussian wave function develops a large number of nodes
and so the wave function even at modest times is of the form
Gaussian time a high order polynomial.  In spite of this,
$\langle x^2(t) \rangle$ shows rather simple behavior.  It
oscillates with a constant amplitude for a reasonable period of
time with an envelope that oscillates with a much longer time
constant which increases with $N$.  The Hartree and leading order
large-$N$ approximations just oscillate with fixed amplitude. The
NLOLN blows up in this regime.  BVA attempts to track the
contraction of the envelope but then contracts to a fixed point.
The DDSA violates energy conservation at order $1/N$ so it becomes
numerically inaccurate when $1/N$ effects become important which
is at a time $t \propto N$.  Both BVA and DDSA do however stay
bounded and positive definite during the time period of our
numerical simulations. Higher order correlation functions show
more complicated behavior and the approximations presented here
are only accurate for a few oscillations in the regime $3 < N <
20 $ consistent with the increasingly complicated evolving
structure of the wave function.

In Figs.~\ref{SD.fig:n3_x2} through \ref{SD.fig:n21_x2}, we show the
results for $\langle x^2(t) \rangle$ as a function of $t$, comparing
the bare vertex, the dynamic Debye screening, and the large-$N$
approximations to the exact solution, for $N=3$, $10$, and $21$.
In Figs.~\ref{SD.fig:n3_chi} to \ref{SD.fig:n21_chi}, we show the same
results for $\langle \chi(t) \rangle$ as a function of $t$, and in
Figs.~\ref{SD.fig:n3_chi2} through \ref{SD.fig:n21_chi2}, we give the
results for $\langle \chi^2(t)$ [For detailed views of these figures
in color, see our web site at: http://www.theory.unh.edu/resum].

In our previous studies\cite{ref:paper1} of the large-$N$
approximation, we found that the next to leading order large-$N$
approximation had the feature that the effective potential was not
defined at small $x$ for $N \leq 20$, for our parameter set, and it
was not until $N$ was greater than about $20$ that the large-$N$
expansion produced bounded values for $\langle x^2(t) \rangle$.  This
result is reproduced here.  For the limit $N \rightarrow \infty$ the
quantity $\langle x^2(t) \rangle$ corresponds to harmonic
oscillations.  At finite $N$, however, the exact solution for $N \ge
21$ has the property that the envelope of these oscillations
contracts.  As noted in the figures, only the bare vertex
approximation attempts to follow this contraction.  At $N = 21$, the
BVA is accurate up to a $t \approx 130$ before overshooting and then
oscillating about a fixed point.  This fixed point behavior shows that
this approximation still neglects some important quantum phase
information present in the exact solution.

In contrast to the NLOLN approximation, which breaks down for $N <
21$, both the BVA and the DDSA have the feature that $\langle
x^2(t) \rangle$ remains positive definite, as well as being
bounded at all $N$.  This is true for all the expectation values
that contribute to the energy.  This conclusion is purely based
on numerical evidence. We do not have a proof that this
approximation corresponds to a positive definite probability
distribution.  However, all the moments we have studied (a total
of five, as shown in Fig.~\ref{fig:energies}), are all bounded.

The DDSA is more accurate than the second order large-$N$
approximation for $N$ less that $20$, but for $N$ greater than $20$,
the reverse becomes true.  However, neither approximation captures the
true nonlinear shrinking of the envelope of the oscillations, even for
$N$ greater than $20$.

Energy is conserved for the bare vertex and the second order large-$N$
approximations, but not for the dynamic Debye screening
approximations, as pointed out in section~\ref{sec:DDSA}.  This is a
serious drawback to the dynamic Debye screening approximation.

In all these figures, one can see that the bare vertex approximation
tries to follow the envelope of the exact curve, whereas the dynamic
Debye screening approximation does not do so.  This is particularly
striking for the cases when $N$ is less than $21$, where the dynamic
Debye screening approximation yield unphysically large values for the
expectation values.

In the BVA approximation we observe that $\langle x^2(t) \rangle$
at late times has an envelope of decreasing oscillations about a
fixed point.  In fact as seen in Fig.~\ref{fig:energies} all the
contributions to the energy in the BVA have the same feature that
they asymptote to a fixed point. In Fig.~\ref{fig:energies} we
display all five contributions to the energy at $N=10$ to
demonstrate this fact. In contrast, as seen in the very long time
run shown in Fig.~\ref{SD.fig:n21_chi2}, the exact solutions
exhibit ``recurrence'' patterns of motion which are not captured
in the BVA. In the $1+1$ dimensional field theory simulations of
ref.~\cite{ref:berges}, all the Fourier components of the two
particle correlation function showed this behavior which was
given as evidence for thermalization.  So one hopes that this
``defect'' of the BVA approximation in a quantum mechanics
setting, will instead have the correct physics of thermalization
in a field theory application where Poincar\'e recurrence times
are expected to become very large.  To see if this is true, we
intend to study the BVA in classical 1+1 dimensional field theory
where again exact simulations can be performed\cite{abw}.

In summary we have found that both resummation methods described
here, the BVA and the DDSA, produce positive definite and
apparently bounded results for expectation values at all values
of $N$.  The bare vertex approximation appears to provide the
best description of the motion, but cannot describe recurrences
of the motion.  Still, it provides an energy conserving and
reasonably accurate description, and is a dramatic improvement
over the next to leading order large-$N$ approximation when $N <
N_{\text{crit}} = 21$.  As mentioned earlier, in the single
particle quantum mechanics problem we studied here, the graphs do
not correspond to particle collisions, so there is no possibility
of studying thermalization.  Thermalization questions need to be
addressed in field theory applications.  It will be important to
show that the BVA approximation will lead to thermalization of
arbitrary initial data as found in the 3-loop approximation of
ref.~\cite{ref:berges} when applied to 1+1 dimensional quantum
field theory.  We would also like to study the analogue of the BVA
approximation for a gaussian ensemble of initial conditions for a
1+1 dimensional classical field theory since that can also be
studied exactly numerically\cite{abw}.  These authors have shown
that the classical field theory indeed thermalizes and we would
like to know how accurately the classical version of our
approximation captures this physics. This will be the subject of
a future publication.
\bigskip

%
%
\begin{acknowledgments}

We wish to thank Salman Habib for helpful discussions on
understanding the numerical simulations and for continued
advice.  We wish to thank Prof.~Gabor Kalman for explaining
relevant plasma conductivity approximations, and Emil Mottola for
suggesting our study of the ``dynamic Debye screening''
approximation and explaining its derivation from the CJT
formalism.  We also thank Juergen Berges for discussing with us
his recent results on thermalization in a related approximation.
JFD is supported in part by the U.S.~Department of Energy under
grant DE-FG02-88ER40410.  He would like to thank the T-8 theory
group at LANL, and the Institute for Nuclear Theory at the
University of Washington, for hospitality during the course of
this work. FC would like to thank Boston College and UNH for
hospitality during the course of this work.

\end{acknowledgments}

%
%

%
%
\end{document}